\documentclass[12pt]{article}
\usepackage{color}
\usepackage[dvips]{graphicx}
\usepackage{dcolumn}
\usepackage{amsmath}
\usepackage{url}

\begin{document}
\Large
\begin{center}
  Diagrammatic Multiplet-Sum Method (MSM) Density-Functional Theory (DFT):
  II. Completion of the Two-Orbital Two-Electron Model (TOTEM) with 
  an Application to the Avoided Crossing in Lithium Hydride (LiH)
\end{center}
\normalsize

\vspace{0.5cm}

\noindent
Mark E.\ CASIDA\\
{\em Laboratoire de Spectrom\'etrie, Interactions et Chimie th\'eorique 
(SITh),
D\'epartement de Chimie Mol\'eculaire (DCM, UMR CNRS/UGA 5250),
Institut de Chimie Mol\'eculaire de Grenoble (ICMG, FR2607), 
Universit\'e Grenoble Alpes (UGA)
301 rue de la Chimie, BP 53, F-38041 Grenoble Cedex 9, FRANCE\\
e-mail: mark.casida@univ-grenoble-alpes.fr} 

\vspace{0.5cm}

\noindent
Abraham PONRA\\
{\em Department of Physics, Faculty of Science, 
University of Maroua, P.O.\ Box 814, Maroua, CAMEROON\\
e-mail: abraponra@yahoo.com}

\vspace{0.5cm}

\noindent
Carolyne BAKASA\\
{\em Technical University of Kenya, P.O.\ Box 52428-00200,
Haile Selassie Avenue, Nairobe, KENYA\\
e-mail: carolyne.bakasa@gmail.com}

\vspace{0.5cm}

\noindent
Anne Justine ETINDELE\\
{\em Higher Teachers Training College, University of Yaounde I,
P.O.\ Box 47, Yaounde, CAMEROON\\
e-mail: anne.etindele@univ-yaounde1.cm}

%

\vspace{0.5cm}

\begin{center}
{\bf Abstract}
\end{center}

The Ziegler-Rauk-Baerends multiplet sum method (MSM) assumes that 
density-functional theory (DFT) provides a good description of states
dominated by a single determinant.  It then uses symmetry
to add static correlation to DFT.  In our previous article (Article I)
[{\em J.\ Chem.\ Phys.}\ {\bf 159}, 244306 (2023)], we introduced
diagrammatic MSM-DFT as a tool to aid in extending MSM-DFT to include
the nondynamic correlation needed for making and breaking bonds {\em even 
in the absence of symmetry}.  An attractive feature of this approach is
that no functional-dependent parameters need to be introduced, though
choices are needed in making correspondances between wave function
theory (WFT) and MSM-DFT diagrams.  The preliminary examples in Article I
used the two-orbital two-electron model (TOTEM) less completely than
could have been the case as we wanted to limit calculations to diagonalizing
$2 \times 2$ matrices, which can be done by solving a simple quadratic
equation.  Diagrammatic MSM-DFT is extended here to treat
the full TOTEM and it is shown that the {\em unsymmetric} lithium hydride (LiH)
molecule dissociates into neutral atoms when diagrammatic MSM-DFT techniques
are used to introduce a proper description of the avoided crossing between 
ionic bonding and covalent bonding states.  This involves diagonalizing
a $3 \times 3$ matrix which requires going beyond solving a quadratic
equation but is still trivial these days. The method is
tested for Hartree-Fock and for three functionals (LDA, PW91, and B3LYP).
All the functionals yield similar results as should be expected for a 
properly-formulated parameter-free theory.  Agreement with available
estimates show that the magnitude of the coupling element introduced
here is excellent.  However more work will be needed to obtain
quantitative agreement between our diagrammatic MSM-DFT ground-state
potential energy curve and that found from high-quality {\em ab initio} 
calculations.


\section{Introduction}
\label{sec:intro}

\marginpar{\color{blue} TD, DFT}
Time-dependent (TD) density-functional theory (DFT) is one of the most
widely used methods for studying electronic spectra and
photochemistry of medium- to large-sized molecules, but it faces 
a number of challenges \cite{C09,CH12}.
In particular, photochemical applications frequently encounter the
problem of triplet instabilities (where a TD-DFT triplet excitation
\marginpar{\color{blue} LUMO, HOMO}
energy becomes imaginary) or spin-wave instabilities [where the energy
of the lowest unoccupied molecular orbital (LUMO) falls below the
energy of the highest occupied molecular orbital (HOMO)].  Both types of
instabilities arise from and contain information about deficiencies in the ground-state
\marginpar{\color{blue} GS, SODS, DODS}
(GS) wave function.  For example, a triplet instability in a spin-restricted
[same orbitals for different spins (SODS)] calculation indicates that there
is a lower energy symmetry-broken spin-unrestricted solution 
[different orbitals for different spins (DODS)] \cite{CGG+00} while the 
spin-wave instability indicates an effective break down of noninteracting
\marginpar{\color{blue} NVR, DFA}
$v$-representability (NVR) \cite{CH12}.  Such problems are not entirely
unexpected because density-functional approximations (DFAs) are designed
to include dynamic correlation but not strong correlation effects such
as static and nondynamic correlation.  Following the classification 
scheme of Bartlett and Stanton \cite{BS94}, dynamic correlation is that
which is present when a single-determinant (SDET) is an adequate first
approximation, static (or degenerate) correlation is present due to
degeneracies which usually those found from symmetry, while nondynamic
\marginpar{\color{blue} SDET, MDET}
(or quasidegerate) correlation arises due to near degeneracies such as
arise near transition states in chemical reactions when bonds are being
made or broken.  In these latter two cases, a multideterminantal (MDET)
wave function is expected to be a better first approximation and absolutely
essential for exploring ground- and excited-state potential energy surfaces
\marginpar{\color{blue} PES, PEC}
(PESs) and potential energy curves (PECs).   While our
ultimate objective is a TD-MDET-DFT for photochemical dynamics simulations,
this article is {\em not} about TD-DFT.  Instead, this article forcuses
on exploring one of the simplest MDET extensions of DFT, namely the
\marginpar{\color{blue} MSM, WFT}
multiplet-sum method (MSM) \cite{ZRB77,D94,PEMC21}.  As the name suggests, 
MSM-DFT was designed to treat static correlation by making extensive use 
of symmetry arguments.  Additional ``symmetry-free'' arguments are needed
to include nondynamic correlation.  These may be obtained---or more exactly
guessed within certain logical constraints---by finding symmetries between 
DFT and wave function theory (WFT) using the diagrammatic techniques that 
we have described in an earlier article \cite{PBEC23}, hereafter known as 
\marginpar{\color{blue} Article {\bf I}}
Article {\bf I}.  However it was unable to account for the avoided crossing in 
the PEC of lithium hydride, LiH.  Here we explore a
further generalization of diagrammatic MSM-DFT suitable for a {\em general}
\marginpar{\color{blue} TOTEM}
two-orbital two-electron model (TOTEM) problem and we will test how well it
works for the PEC of LiH.

Diagrammatic MSM-DFT
is a ``simple'' tool for analysing MSM-DFT and comparing it with
\marginpar{\color{blue} CI, CIME}
WFT configuration interaction (CI) matrix element (CIME)
diagrams.  CIME diagrams are described in chapter 4 of Shavitt and Bartlett's 
\marginpar{\color{blue} SI}
well-known book \cite{SB09} and in the Supplementary Information (SI) 
of Article {\bf I}.  They are simpler than 
ordinary Feynman diagrams and are referred to in Shavitt and Barlett's book
as ``the nonstandard notation.''  As explained in Article {\bf I}, our goal is to 
construct a small CI MSM-DFT matrix (the smallest that still includes the basic
physics of our problem) whose elements follow the 
\marginpar{\color{blue} FPP}
\begin{quotation}
\noindent
{\em Fundamental Pragmatic Principle} (FPP):  Whenever possible, all CI MSM-DFT
matrix elements must be obtainable from SDET calculations.
\end{quotation}
For consistency reasons, we also require that each CI MSM-DFT matrix element
\marginpar{\color{blue} xc} 
reduce to ordinary CIME diagrams when the exchange-correlation (xc) 
functional is expanded to second order
\marginpar{\color{blue} EXAN} 
under the exchange-only ansatz (EXAN),
\begin{eqnarray}
  \langle p \vert v_{xc}^\sigma \vert q \rangle & \rightarrow & 
  \langle p \vert {\hat \Sigma}_x^\sigma \vert q \rangle 
  = -\sum_{k^\sigma} (pk^\sigma \vert f_H \vert k^\sigma q) \, , \nonumber \\
  (pq \vert f_{xc}^{\uparrow,\uparrow} \vert rs) & \rightarrow &
  -(ps \vert f_H \vert rq) 
  \mbox{\,\,\,\,\, , \,\,\,\,\,}
  (pq \vert f_{xc}^{\uparrow,\downarrow} \vert rs )  \rightarrow  0 \, ,
  \label{eq:intro.1}
\end{eqnarray}
where ${\hat \Sigma}_x^\sigma$ is the spin $\sigma$ exchange operator
(often represented by $-{\hat K}^\sigma$ in the quantum chemistry literature),
$k^\sigma$ refers to spin $\sigma$ occupied orbitals, and
\begin{eqnarray}
  (pq \vert f \vert rs) & = & \int \int \psi_p^*(1) \psi_q(1) f(1,2)
  \psi_r^*(2) \psi_s(2) \, d1 d2 \, \nonumber \\
  f_H^{\sigma,\tau}(\vec{r}_1,\vec{r}_2) & = & 
   \frac{\delta_{\sigma,\tau}}{r_{1,2}} 
  \mbox{\,\,\,\,\, , \,\,\,\,\,}
  f_{xc}^{\sigma,\tau}(\vec{r}_1,\vec{r}_2)  =  
     \frac{\delta^2 E_{xc}[\rho_\uparrow, \rho_\downarrow]}
                    {\delta \rho_\sigma(\vec{r}_1) \delta  \rho_\tau(\vec{r}_2)}
  \, ,
  \label{eq:intro.2}
\end{eqnarray}
where $\sigma$ and $\tau$ are respectively the spins of electrons 1 and 2.
These formulae assume either a Hartree-Fock (HF) or pure DFT [local (spin) 
\marginpar{\color{blue} HF, LDA\\GGA} 
density approximation (LDA) or generalized gradient approximation (GGA)] form 
of the DFT xc functional but are easily extended to 
hybrid functionals and meta GGAs.  

While we might expect to have to abandon the FPP once the number of 
SDET energies becomes inconveniently large, it will be shown in the
present work that this is not the case for the TOTEM.  However, we 
go beyond the orginal 
Ziegler-Rauk-Baerends-Daul approach in that we will make use of the 
Kohn-Sham DFT hamiltonian (in its HF, LDA, GGA, and etc.\ form) and 
not just total energies.  Diagonal elements of our CI MSM-DFT matrix 
are SDET DFT energies.  Some of the off-diagonal elements are obtained 
using symmetry arguments exactly as is done in the traditional MSM-DFT 
procedure for capturing static correlation.  

In cases where symmetry arguments are lacking for determining an off-diagonal 
element of the CI MSM-DFT matrix, then we must guess what DFT expression to 
use.  This is tricky.  For example, in Article {\bf I}, it was emphasized and
shown by example that it can be disasterous to just replace missing CI matrix
elements with corresponding formulae from WFT.  However,
if the CIME of the undetermined CI MSM-DFT matrix is identical (for 
real orbitals) to the CIME of a known CI MSM-DFT matrix element, then we 
may just use the known CI MSM-DFT matrix element in the place of the unknown
CI MSM-DFT matrix element.  As shown in Article {\bf I}, this was sufficient 
to create a method where the hydrogen molecule, H$_2$, ground state 
dissociates correctly without symmetry breaking.  (Article {\bf I} also used
the diagrammatic approach to produce an alternative way to calculate
a CI MSM-DFT matrix element for O$_2$.)  

Here we will make 
further use of analogies between the CIME diagrams and diagrammatic 
MSM-DFT diagrams to fill in the rest of the CI MSM-DFT matrix in the
TOTEM for LiH where there is no spatial 
symmetry to help us.  {\em The fact that we are generalizing the original
heavily symmetry-dependent MSM-DFT to cases without spatial symmetry
is very significant as our fundamental objective is to get away from
symmetry constraints whenever this is possible.}

MDET theories typically divide orbitals into three sets---namely
occupied, active, and unoccupied.  Full CI within the active space is 
\marginpar{\color{blue} CAS($n$,$m$)} 
referred to as a CAS($n$,$m$) treatment of $n$ electrons in $m$
orbitals.  CAS(2,2) or the TOTEM
is particularly important because the lines drawn by chemists in 
\marginpar{\color{blue} VB} 
molecular structure diagrams represent electron pair bonds which 
valence-bond (VB) theory tells us requires a MDET, rather than a SDET,
treatment for describing bond breaking and bond formation.  Whether
drawn as lines or pairs of dots, and even when the lone pairs are
\marginpar{\color{blue} LDS} 
implicit, we call these Lewis dot structures (LDSs).

To a first approximation, bonding in a diatomic occurs between an electron 
\marginpar{\color{blue} AO, MO} 
in an atomic orbital (AO) on one atom and an electron in an AO on the
other atom to create two molcular orbitals (MOs)---namely the 
HOMO (H) and the LUMO (L).
\marginpar{\color{blue} H, $i$,\\ L, $a$} 
We will refer to the HOMO as orbital $i$ and to the LUMO
as orbital $a$.

Within this TOTEM, there are only four determinants with spin quantum 
number $M_S=0$.  We will write these in second-quantized form as 
$\vert \Phi \rangle = \vert i \bar{i} \rangle$, 
$\vert \Phi_i^a \rangle = a^\dagger i \vert \Phi \rangle = \vert a \bar{i} \rangle$,
$\vert \Phi_{\bar a}^{\bar i} \rangle = {\bar a}^\dagger {\bar i} \vert \Phi \rangle
= \vert i {\bar a} \rangle$, and $\vert \Phi_{i,{\bar i}}^{a,{\bar a}} \rangle
= a^\dagger i {\bar a}^\dagger {\bar i} \vert \Phi \rangle = \vert a {\bar a} \rangle$.
Using this order to label the columns and rows, our CI matrix is
\begin{equation}
  {\bf H} = \left[ \begin{array}{cccc} 
            E_0 & D & D & B \\
             D  & E_M & A & C \\
             D  & A   & E_M & C \\
             B  & C   & C & E_D 
            \end{array} \right] \, .
  \label{eq:intro.3}
\end{equation}
The solutions of the CI equation ${\bf H} \vec{C} = E \vec{C}$ must be eigenfunctions
of ${\hat S}^2$.  This is accomplished by transforming the original SDET basis into
spin-adapted configurations, namely the triplet
\begin{equation}
   \vert \Phi_T \rangle =  
   \frac{1}{\sqrt{2}} \left( \vert \Phi_{\bar i}^{\bar a} \rangle
                    - \vert \Phi_i^a \rangle \right) 
  \label{eq:intro.4}
\end{equation}
and the three singlet configuration state functions,
\begin{equation}
   \vert \Phi_0 \rangle   =   \vert \Phi \rangle
   \mbox{\,\,\,\,\, , \,\,\,\,\,}
   \vert \Phi_S \rangle  =  \frac{1}{\sqrt{2}} \left( \vert \Phi_{\bar i}^{\bar a} \rangle
                    + \vert \Phi_i^a \rangle \right) 
  \mbox{\,\,\,\,\, , \,\,\,\,\,}
  \vert \Phi_D \rangle  =   \vert \Phi_{i {\bar i}}^{a {\bar a}} \rangle 
  \, .
  \label{eq:intro.5}
\end{equation}
Having constructed $\hat{S}^2$ eigenfunctions, then we can transform the 
CI matrix to this new basis to give us a single triplet energy,
\begin{equation}
  E_T = E_M - A \, . 
  \label{eq:intro.6}
\end{equation}
and a $3 \times 3$ singlet CI matrix,
\begin{equation}
  {\bf H}_S = \left[ \begin{array}{ccc}
  E_0 & \sqrt{2} D & B \\
  \sqrt{2} D & E_M+A & \sqrt{2} C \\
  B & \sqrt{2}C & E_D 
  \end{array} \right] \, .
  \label{eq:intro.7}
\end{equation}
\marginpar{\color{blue} M}
Here M refers to the mixed symmetry states,
\begin{equation}
  \vert \Phi_M \rangle = \vert \Phi_i^a \rangle 
  \mbox{\,\,\,\,\, , \,\,\,\,\,}
  \vert \Phi_{\bar M} \rangle =  \vert \Phi_{\bar i}^{\bar a} \rangle 
  \mbox{\,\,\,\,\, , \,\,\,\,\,}
  E_M = E[\Phi_M] = E[\Phi_{\bar M}] \, .
  \label{eq:intro.8}
\end{equation}
In practice, Eq.~(\ref{eq:intro.6}) is solved for $A$ which is thus used in 
Eq.~(\ref{eq:intro.7}).
The squares of the coefficients of the eigenvectors $\vert \Psi \rangle
= C_0 \vert \Phi_0 \rangle + C_S \vert \Phi_S \rangle + C_D \vert \Phi_D \rangle$ 
provide the weights of the three different contributions to the total
wave function.

In Article {\bf I}, we neglected both $C$ and $D$.  This is justified by symmetry
in the case of H$_2$, but is no longer justifiable for LiH.  All terms 
$A$, $B$, $C$, and $D$ will be included here.  One price that we shall 
have to pay is the need to diagonalize a $3 \times 3$ matrix 
but this is hardly a problem these days as it can be done with a 
good calculator or, in the present paper, with very simple {\sc Python} 
programs and we shall see that even this $3 \times 3$ matrix eigenvalue problem
may be reduced to a $2 \times 2$ matrix eigenvalue problem for LiH
when using an ensemble reference.

Let us now focus on the special case of LiH.  Elementary textbooks treat
the $\sigma$ bond in LiH as mainly due to the overlap of the $2s$ AO
on Li, $s_{\mbox{Li}}$, with the $1s$ AO on H, $s_{\mbox{H}}$.  As the
$s_{\mbox{H}}$ AO is lower in energy than the $s_{\mbox{Li}}$ AO, the 
HOMO is dominated by the $s_{\mbox{H}}$ AO while the LUMO is dominated by
the $s_{\mbox{Li}}$ AO.  Orthonormality considerations then give,
\begin{eqnarray}
  i & = &  {\cal N}_1 \left( s_{\mbox{H}} + \eta_1 s_{\mbox{Li}} \right)
  \mbox{\,\,\,\,\, , \,\,\,\,\,}
  a = {\cal N}_2 \left( s_{\mbox{Li}} - \eta_2 s_{\mbox{H}} \right)
  \, , \nonumber \\
  {\cal N}_1 & = & 1/\sqrt{1+2S \eta_1 + \eta_1^2} 
  \mbox{\,\,\,\,\, , \,\,\,\,\,}
  {\cal N}_2 = 1/\sqrt{1-2S \eta_2 + \eta_2^2} \, , \nonumber \\
  S & = & \langle s_{\mbox{Li}} \vert s_{\mbox{H}} \rangle 
  \mbox{\,\,\,\,\, , \,\,\,\,\,}
  \eta_2 = \frac{\eta_1+S}{1+S\eta_1} \, ,
  \label{eq:intro.9}
\end{eqnarray}
at the equilibrium geometry where the electronic configuration is
\begin{equation}
  [1\sigma(\mbox{Li $1s$})]^2 [2\sigma(\mbox{H $1s$})]^2
  [3\sigma(\mbox{Li $2s$})]^0 [1\pi(\mbox{Li $2p$})]^0 \cdots \, .
  \label{eq:intro.10}
\end{equation}
In VB terms, this corresponds to the ionic LDS
$\left[ \mbox{Li$^+$} \,\,\,\, \mbox{H:$^-$} \right]$.

\begin{figure}
\begin{center}
\includegraphics[width=0.6\textwidth]{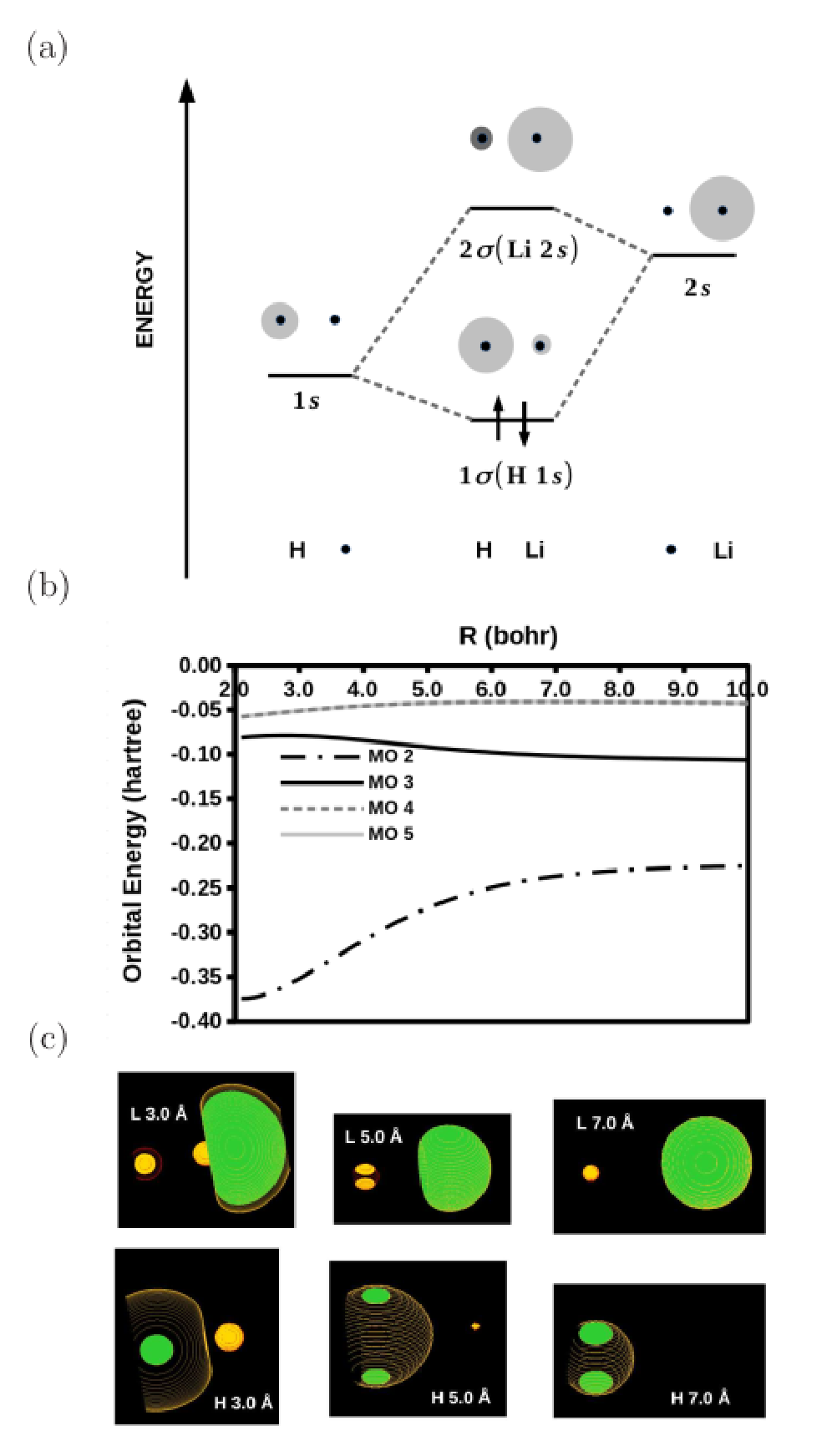}
\end{center}
\caption{
Reference (half-occupied) LDA frontier MOs (c) and their energies (b) at
various distances. MO visualization was done with {\sc Molden} \cite{molden}.
As expected from textbook MO theory (a), the HOMO is mainly
on H (left atom) while the LUMO is mainly on Li (right atom).
\label{fig:LiH_LDA_MOs}
}
\end{figure}

A second feature of MDET calculations is that we
must choose a reference.  Our reference is obtained by an equally-weighted
ensemble of all four $M_S=0$ states whose two-electron density matrix 
operator is,
\begin{equation}
  \hat{\Gamma} = \frac{1}{4} \left( \vert \Phi_0 \rangle \langle \Phi_0 \vert
  + \vert \Phi_S \rangle \langle \Phi_S \vert 
  + \vert \Phi_D \rangle \langle \Phi_D \vert
  + \vert \Phi_T \rangle \langle \Phi_T \vert \right) \, ,
  \label{eq:intro.11}
\end{equation}
for which $\langle \hat{S}^2 \rangle = 1/2$ and the corresponding one-electron
reduced density operator,
\begin{equation}
  \hat{\gamma} = \frac{1}{4} \left( \vert i \rangle \langle i \vert
  + \vert {\bar i} \rangle \langle {\bar i} \vert 
  + \vert a \rangle \langle a \vert
  + \vert {\bar a} \rangle \langle {\bar a} \vert \right) \, ,
  \label{eq:intro.12}
\end{equation}
with half an electron of each spin in the HOMO and half an electron of each
spin in the LUMO.  Our calculation with this ensemble reference (see
Sec.~\ref{sec:details} for computational details) show that this leads
to easy convergence of a SODS solution without any significant symmetry
\marginpar{\color{blue} VWN} 
breaking.  However it is important to keep in mind that we are using orbitals
which are not specifically optimized for the GS but which rather are an
attempt to treat all the states in an impartial manner.
LDA MOs with the Vosko-Wilk-Nusair
(VWN) parameterization \cite{VWN80} of Ceperley and Alder's quantuim 
Monte Carlo results for the electron gas \cite{CA80} are shown 
in {\bf Fig.~\ref{fig:LiH_LDA_MOs}}.
Similar results were found with other DFAs and are shown explicitly in the 
SI associated with this article.
In all cases, the $1\sigma$ and $2\sigma$ MOs remain bound at all 
\marginpar{\color{blue} $R$} 
bond distances, $R$.  
Notice how the $1\sigma$ and $2\sigma$ approach each other at large $R$
but that they never become degenerate.  Some hybrid $sp$ character is also
present and, notably, evident in the figure in the $2\sigma$ orbital at 
$R = \mbox{3.0 {\AA}}$.

\begin{figure}
\begin{center}
\includegraphics[width=0.8\textwidth]{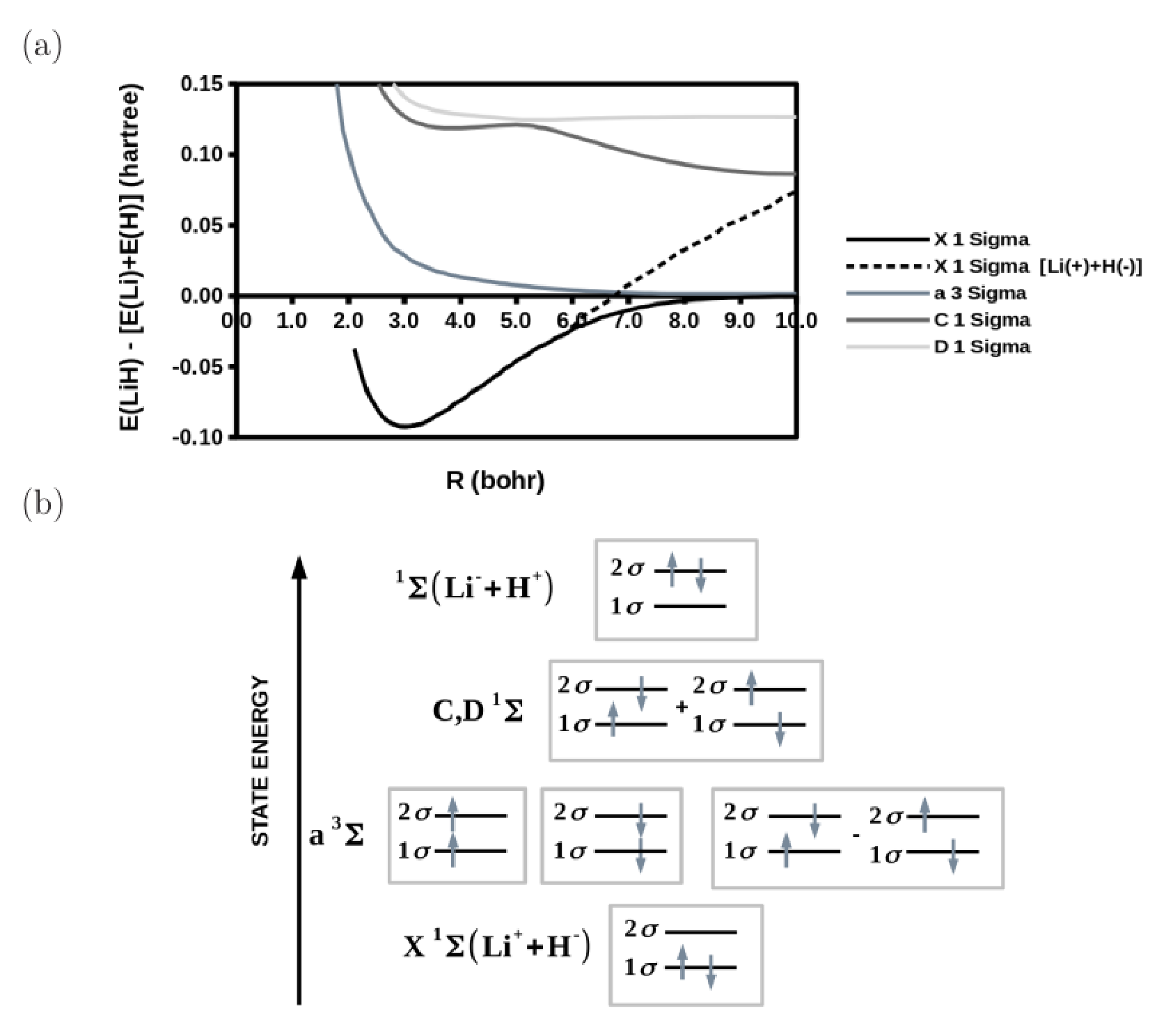}
\end{center}
\caption{
(a) LiH molecule high-quality (``EXACT'') PECs digitized from graphs in 
Ref.~~\cite{GL06}.  (b) PEC assignments.  Reproduced from the SI in Article {\bf I}.
\label{fig:LiHEXACT}
}
\end{figure}

\marginpar{\color{blue} EXACT} 
{\bf Figure~\ref{fig:LiHEXACT}} shows high-quality (which we shall refer to as
EXACT) curves against which we will compare our calculations.  A rough 
assignment has also been included in part (b) of the diagram.  Note that the 
$X$ $^1\Sigma(\mbox{Li$^+$+H$^-$})$ curve was obtained by hand tracing
through avoided crossings.  Also note that the exact nature of the 
$C \,^1\Sigma$ and $D \,^1\Sigma$ curves is complicated by an obvious
avoided crossing with a state not present in our simple model.
Most important for the present work is that the initial [Li$^+$  H:$^-$]
state would dissociate incorrectly following the dotted diabatic line.
Mixing with some other state leads to an avoided crossing which leads
to the correct [Li$\uparrow$ + H$\downarrow$ $\leftrightarrow$ Li$\downarrow$
+ H$\uparrow$] gas-phase dissociation.  This was missing in Article {\bf I}
because of our neglect of the $C$ and $D$ matrix elements in 
Eq.~(\ref{eq:intro.7}).

Since our goal is to extend the MSM-DFT method, let us have a brief look
at what has been done with (TD)-DFT for LiH.  Of course, LiH is very small
and, as such, is primarily a toy for us for testing out ideas.  The few
references that we have found that focused on both DFT and LiH have 
also used LiH for this purpose.
We have also included some routine DFT calculations of 
our own, labeled ``relaxed'' in {\bf Table~\ref{tab:MSM_BE}}.  These
show that the overall equilibrium bond length is quite reasonable when
calculated either at
\marginpar{\color{blue} FOCK} 
the HF or at the DFT level.  HF (labeled FOCK in the table) severely
underestimates the bond energy, but the various DFAs do quite well compared
with the EXACT result.
\begin{table}
\begin{center}
\begin{tabular}{ccc}
\hline \hline
Method & Bond Length & Bond Energy \\
\hline
EXACT$^a$             & 3.042 bohr  & 0.09278 Ha  \\
\multicolumn{3}{c}{FOCK} \\
relaxed               & 3.029 bohr  & 0.05497 Ha \\
unrelaxed             & 2.991 bohr  & 0.03227 Ha \\
unsymmetrized         & 3.114 bohr  & 0.05863 Ha  \\
symmetrized           & 3.114 bohr  & 0.05863 Ha  \\
from $\Phi_D$         & 3.114 bohr  & 0.05863 Ha  \\
from $\Phi_0$         & 3.114 bohr  & 0.05863 Ha  \\
\multicolumn{3}{c}{VWN} \\
relaxed               & 3.030 bohr  & 0.09781 Ha \\
unrelaxed             & 2.963 bohr  & 0.04111 Ha \\
unsymmetrized         & 3.213 bohr  & 0.07782 Ha  \\
symmetrized           & 3.207 bohr  & 0.07093 Ha  \\
from $\Phi_D$         & 3.178 bohr  & 0.06379 Ha  \\
from $\Phi_0$         & 3.236 bohr  & 0.07815 Ha  \\
\multicolumn{3}{c}{PW91} \\
relaxed               & 3.028 bohr  & 0.08822 Ha \\
unrelaxed             & 2.959 bohr  & 0.03421 Ha \\
unsymmetrized         & 3.250 bohr  & 0.06785 Ha  \\
symmetrized           & 3.245 bohr  & 0.06008 Ha  \\
from $\Phi_D$         & 3.183 bohr  & 0.05035 Ha  \\
from $\Phi_0$         & 3.286 bohr  & 0.06911 Ha  \\
\multicolumn{3}{c}{B3LYP} \\
relaxed               & 3.000 bohr  & 0.09391 Ha \\
unrelaxed             & 2.935 bohr  & 0.04726 Ha \\
unsymmetrized         & 3.139 bohr  & 0.07574 Ha  \\
symmetrized           & 3.120 bohr  & 0.06924 Ha  \\
from $\Phi_D$         & 3.078 bohr  & 0.06277 Ha  \\
from $\Phi_0$         & 3.150 bohr  & 0.07643 Ha  \\
\hline \hline
\end{tabular}
\end{center}
$^a$ From Ref.~\cite{GL06}.
\caption{Ground-state bond lengths and energies obtained using different
functionals and different methods: ``relaxed'' refers to a normal
ground-state geometry optimization, ``unrelaxed'' is the calculation of the
ground-state energy using the unrelaxed MOs from the reference, and the
rest refer to MSM calculations using different choices of coupling matrix
elements.  With the exception of the ``relaxed'' calculations, all
bond lengths and bond energies were found by a a parabollic fit near the
minimum of the ground-state potential energy curve.  The ``unrelaxed''
bond energy is referenced to the separated atoms while the bond energy of
the MSM calculations is referenced to the triplet energy at 10.0 {\AA}.
\label{tab:MSM_BE}
}
\end{table}

\begin{figure}
\begin{center}
\includegraphics[width=0.7\textwidth]{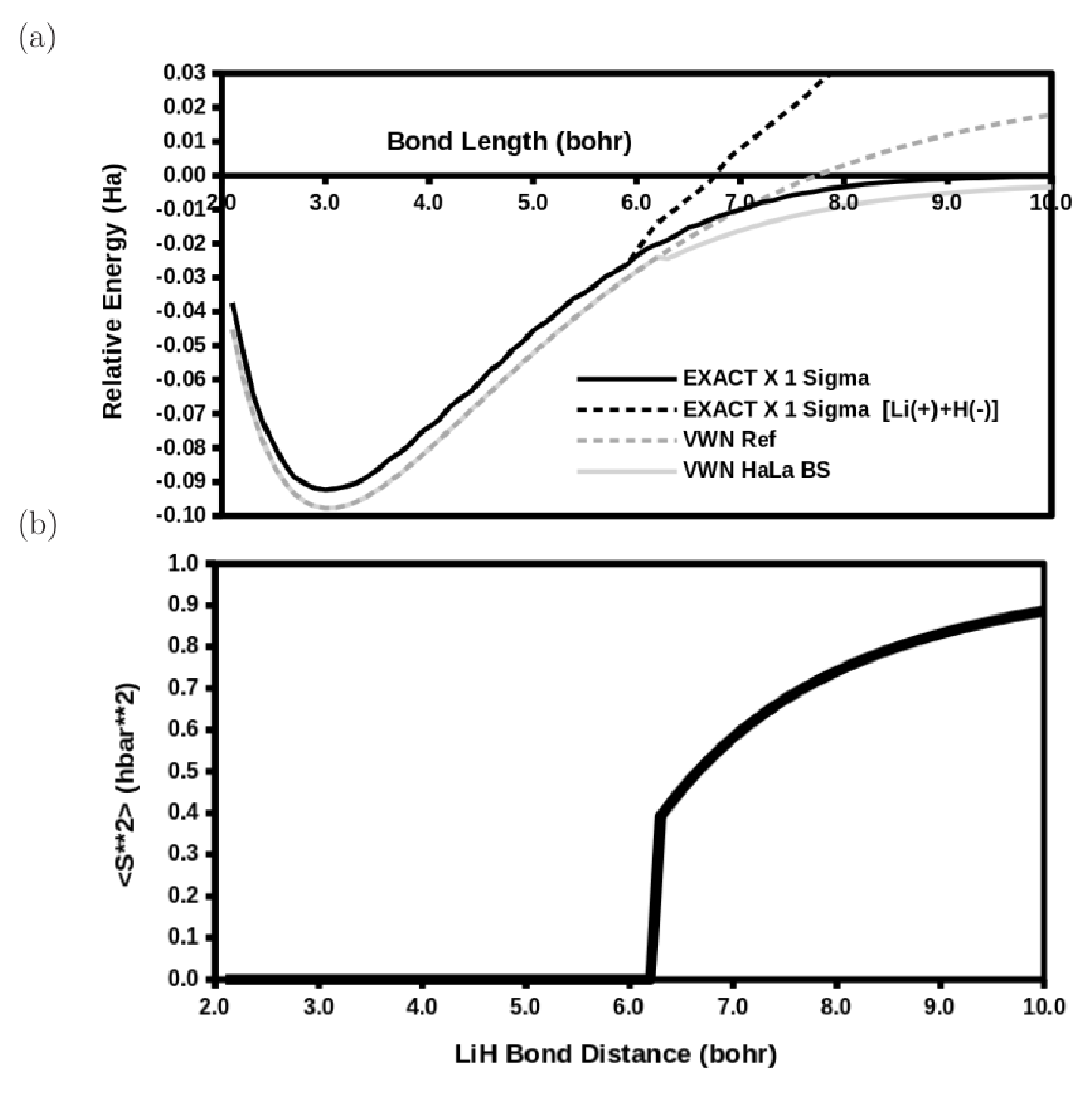}
\end{center}
\caption{
BS VWN calculation of the GS PEC of LiH compared with
the EXACT PEC: (a) PECs relative to the sum of the energies of the separated
neutral atoms calculated at the same level of approximation, 
(b) $\langle {\hat S}^2 \rangle$.
\label{fig:normalLDA}
}
\end{figure}
For reasons of completeness, we would like to mention some other 
(TD)-DFT work on LiH.  Of course, given the small size and relative
``simplicity'' of LiH, we do not expect to find a lot about LiH
in the literature.  But there are some very interesting referenes in the
context of DFT.  The earliest such reference that we have found is in a 
famous paper by Perdew, Parr, Levy, and Balduz on the derivative 
discontinuity in DFT (Fig.~1 of Ref. \cite{PPLB82} and the associated 
discussion) which mentions ``a sudden switch of ground-state character'' 
at a critical distance which is predicted to be at $R_c = \mbox{5.86 bohr}$.
The corresponds roughly to where the [Li$^+$ H:$^-$] dashed diabatic curve
separates from the GS PEC full curve in Fig.~\ref{fig:LiHEXACT}.
\marginpar{\color{blue} PNDD} 
The particle number derivative discontinuity (PNDD) is also responsible 
for the charge transfer problem (briefly reviewed in the SI) that causes 
SODS DFT calculations of dissociating LiH to have fractional charges.  
Conventional DFT calculations of the GS PEC do not look bad at all.
{\bf Figure~\ref{fig:normalLDA}} shows our own VWN
\marginpar{\color{blue} BS} 
calculation showing that a normal {\em broken symmetry} (BS) DFT calculation
does quite well at simulating the EXACT GS PEC provided symmetry.  No
symmetry breaking occurs before $R \approx \mbox{6.3 {\AA}}$ where it
is expected to dissociate to $\langle {\hat S}^2 = 1$.  The fact that
$\langle {\hat S}^2 \rangle < 1$ at large $R$ may be related to the PNDD
and the appearance of fractionally charged atoms at infinite separation.
We would like to emphasize two other problems with BS calculations besides
the triplet instability already mentioned.  These are (i) that convergence
of BS calculations becomes much more difficult beyond, and especially at,
\marginpar{\color{blue} CFP} 
the Coulson-Fischer point (CFP) where the symmetry-broken DODS
solutioin falls lower in energy than the SODS solution 
and (ii) that there may be more than one 
way to break symmetry making it difficult to know how to find the best BS 
solution.

Although not obvious from the figure, the absence of the proper PNDD does 
lead to dissociation into fractionally-charged ions.  This, in turn, 
reflects on the description of the [Li$^+$ H:$^-$]/[Li$\uparrow$ 
H$\downarrow$ $\leftrightarrow$  Li$\downarrow$ H$\uparrow$] avoided crossing
because it indicates an improper treatment of coupling between the two
states.  Kaduc and Van~Voorhis attacked this problem by showing
\marginpar{\color{blue} CDFT}
how the avoided crossing could be treated within constrained DFT (CDFT)
by calculating the coupling element between the diabatic curves
(Fig.~30 of Ref.~\cite{KKV12}).  

\marginpar{\color{blue} CT}
Charge transfer (CT) is a particular problem in TD-DFT where CT excitation
energies can be seriously underestimated.  Casida {\em et al}.\ 
used an analysis reminiscent of the present manuscript 
in order to treat charge-transfer within a TOTEM for H$_2$ and LiH 
\cite{CGG+00}.  We emphasize that Article {\bf I} and the present article is 
much more complete in its formalism and exploration than was Ref.~\cite{CGG+00}.
However an important aspect of the work in Ref.~\cite{CGG+00}
is an explicit TD-DFT calculation for LiH showing how the lowest triplet 
excitation energy goes to zero at the CFP where the symmetry-broken
DODS solution falls lower in energy than the SODS solution.  At larger $R$,
beyond this CFP, the triplet excitation energy actually becomes imaginary 
and there are also problems with the singlet excitation energies.

One might have thought the issue of BS TD-DFT calculations to have been
settled with the earlier work, but 11 years after the publication of
Ref.~\cite{CGG+00}, Fuks, Rubio, and Maitra showed that TD-DFT calculations 
seem to work reasonably well for the excited states of LiH \cite{FRM11},
provided symmetry breaking is allowed.  It makes sense that this could
happen for certain states because, as we have commented upon above, 
the different PECs basically correspond to different excited states of the 
neutral Li atom. Another eight years later, Hait, Rettig, and Head-Gordon 
extended the study of the TD-DFT PEC of LiH beyond the CFP, emphasizing
the difficulties encountered \cite{HRH19}. 
Another decade later, Dar and Maitra
discuss a method to improve oscillator strengths in TD-DFT of LiH in a
method \cite{DM23} reminiscent of dressed TD-DFT \cite{MZCB04}.

%
%

The objective of the present work is to continue investigating how
diagrammatic MSM-DFT can be further developed with the hope that it
will provide a MDET DFT formalism which will solve some of the problems
of TD-DFT coming from an inadequate description of strong correlation
in the GS.  This involves finding a way to complete the MSM-DFT CI matrix
[Eq.~(\ref{eq:intro.7})].
Classic symmetry-based MSM-DFT told us how to calculate the $A$ matrix
element.  In Article {\bf I}, we proposed that the $B$ matrix element should
be identical to the $A$ matrix element in the case of real-valued orbitals
and showed that this led to the correct dissociation of H$_2$.  To get the 
avoided crossing right in LiH between the [Li$^+$  H:$^-$] $\rightarrow$
Li$^+$ + H:$^-$ and  [Li$\uparrow$ H$\downarrow$ $\leftrightarrow$
Li$\downarrow$ H$\uparrow$] $\rightarrow$ Li$\cdot$ + H$\cdot$ PECs, 
we also need a way to calculate the $C$ and $D$ matrix elements consistent
with the FPP.  Different, but closely related, formal choices of $C$
and $D$ are presented in the next section (Sec.~\ref{sec:theory}) and the 
results of these different solutions are presented and discussed in 
Sec.~\ref{sec:results}.  We sum up in Sec.~\ref{sec:conclude}.

\section{Diagrammatic MSM-DFT}
\label{sec:theory}

Much of the intricacies of MSM-DFT, such as the choice of the TOTEM
and the choice of an ensemble reference have already been presented
in Article I and reviewed in the previous section (Sec.~\ref{sec:intro}).
Our goal in this section is to show how to use diagrammatic techniques
to come up with educated guesses for the forms of the $C$ and $D$ matrix
elements.  This will involve a little further review but also, we think,
some insights.

\begin{figure}
\begin{center}
\includegraphics[width=0.7\textwidth]{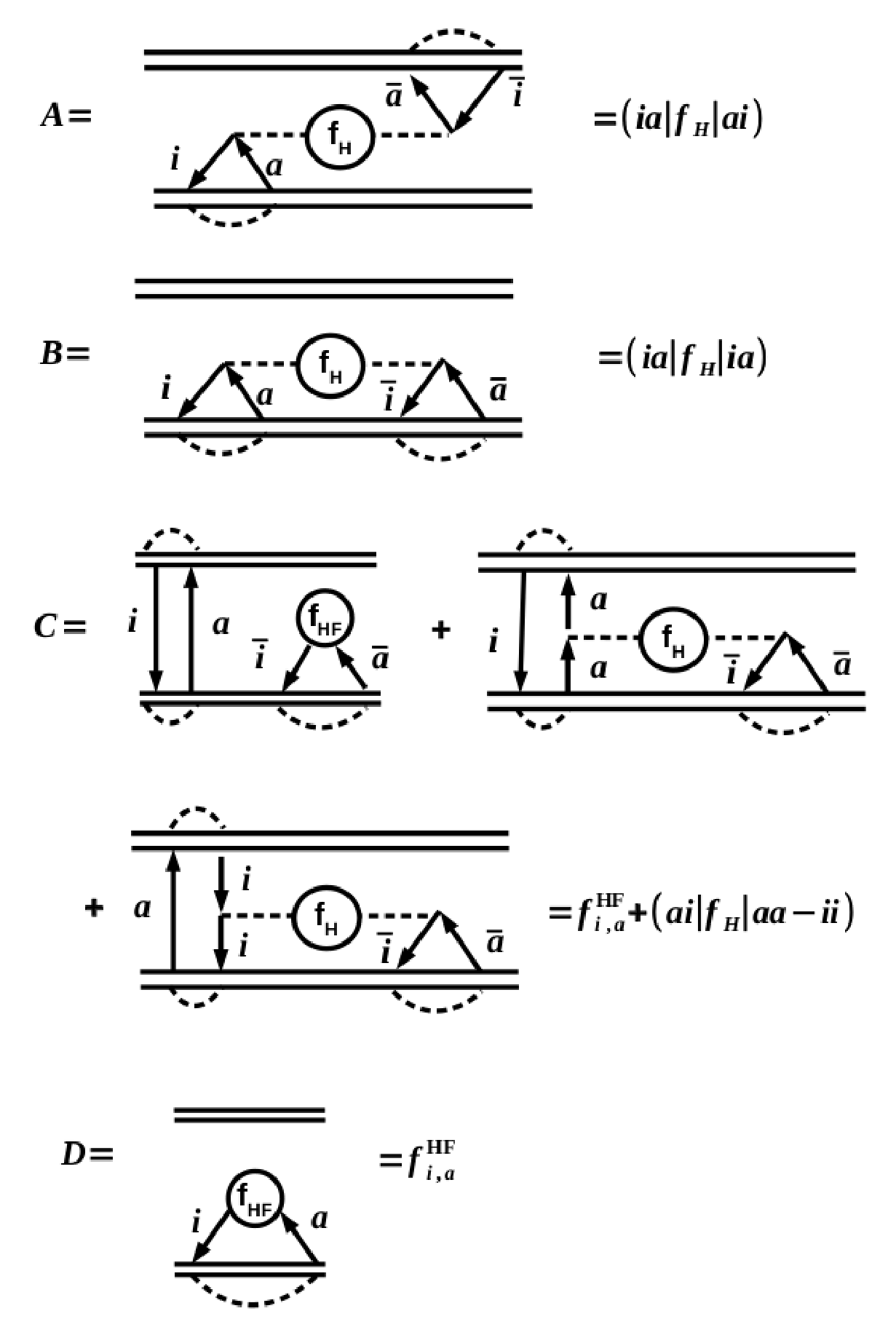}
\end{center}
\caption{
CIME diagrams for the off-diagonal CI matrix elements $A$, $B$, $C$,
and $D$.
\label{fig:CIME}
}
\end{figure}

The WFT CIMEs for the $A$, $B$, $C$, and $D$ terms are shown in 
{\bf Fig.~\ref{fig:CIME}}.  
It is important to realize that these diagrams are valid for any
choice of SDET reference state $\Phi_0$.  This Fermi vacuum need
not be constructed from the canonical MOs.  Hence Brillouin's theorem
($f^{\mbox{HF}}_{i,a}=0$) does not necessarily hold.
Notice how $C$ contains the CT integral $(ia\vert f_H \vert aa-ii)$.
When using canonical MOs, then $C=(ia\vert f_H \vert aa-ii)$ and
$D=0$.  But we will actually be using MOs from our ensemble reference,
constructed from a density matrix $\gamma_{\mbox{ref}}$
with half an electron of each spin in the HOMO and the LUMO.
One consequence of this is that we will also obtain CT contributions 
in the $D$ matrix elements as well as in the $C$ matrix elements 
[Eq.~(\ref{eq:theory.11}) below].  These charge-transfer integrals 
will not be calculated explicitly in the present work but will be 
present implicitly.

The MSM (FPP) requires that we re-express all matrix elements of the CI MSM
matrix in quantities uniquely determined from SDET calculations.  This
is automatic for the diagonal elements: 
$E_0 = \langle \Phi \vert {\hat H} \vert \Phi \rangle$,
$E_T = \langle \Phi_{\bar i}^a \vert {\hat H} \vert \Phi_{\bar i}^a \rangle$,
$E_M = \langle \Phi_i^a \vert {\hat H} \vert \Phi_i^a \rangle$, and
$E_D = \langle \Phi_{i,{\bar i}}^{a,{\bar a}} \vert {\hat H} \vert
\Phi_{i,{\bar i}}^{a,{\bar a}} \rangle$.  It is also possible for the
off-diagonal elements within the HF approximation.  The classic MSM and
spin symmetry argument \cite{ZRB77} gives,
\begin{equation}
  A = E_M - E_T = E[\Phi_i^a] - E[\Phi_{\bar i}^a] \, .
  \label{eq:theory.7}
\end{equation}
For real orbitals,
\begin{equation}
  B = A \, .
  \label{eq:theory.8}
\end{equation}
In Article I, it was shown that this suffices to obtain a reasonable ground state
PEC for H$_2$ without symmetry breaking.  

The key innovation in this article is to make use of off-diagonal elements of
the Fock (or Kohn-Sham) matrix $f_{i,a}[\gamma]$.  Notice that this Fock 
matrix depends only upon the density matrix $\gamma$.  It is thus defined
for both pure-state (SDET) density matrices ($\gamma_0$, $\gamma_M$, 
$\gamma_D$) and for ensemble density matrices $\gamma_{\mbox{ref}}$.
According to Fig.~\ref{fig:CIME},
\begin{equation}
  D = f_{i,a}^{\mbox{HF}}[\gamma] \, ,
  \label{eq:theory.9}
\end{equation}
where $\gamma = \gamma_0$, but is not zero because Brillouin's theorem does 
not hold for our choice of reference MOs. Instead we have that 
$f_{i,a}^{\mbox{HF}}[\gamma_{\mbox{ref}}]=0$
for our ensemble density matrix $\gamma_{\mbox{ref}}$. 
Let us calculate the ensemble Fock matrix from the reduced
density matrix,
\begin{equation}
  \hat{\gamma} = \hat{\gamma}_{\mbox{ref}} 
  - \frac{1}{2} \left(\vert a \rangle \langle a \vert + 
  \vert \bar{a} \rangle \langle \bar{a} \vert \right) + \frac{1}{2}
  \left( \vert i \rangle \langle i \vert 
  + \vert \bar{i} \rangle \langle \bar{i} \vert \right) \, ,
  \label{eq:theory.9.1}
\end{equation}
constructed from the ensemble MOs.  Then
\begin{eqnarray}
  D & = & f^{\mbox{HF}}_{i,a}[\gamma] \nonumber \\
  & = & h_{i,a} + \int \psi_i^*(1) \frac{\rho_{\mbox{ref}}(2)-\rho_a(2)
  + \rho_i(2)}{r_{1,2}} \psi_a(1) \, d1 d2
  \nonumber \\
  & - & \int  \psi_i^*(1)
  \frac{\gamma^\uparrow_{\mbox{ref}}(1,2)-\frac{1}{2}\gamma_a(1,2)+
  \frac{1}{2} \gamma_i(1,2)}{r_{1,2}}
   \psi_a(2) \, d1 d2 \nonumber \\
   & = & f^{\mbox{HF}}_{i,a}[\gamma_{\mbox{ref}}]
   + (ia \vert f_H \vert ii-aa) 
  + \frac{1}{2} (ia \vert \vert aa-ii) \nonumber \\
   & = & f^{\mbox{HF}}_{i,a}[\gamma_{\mbox{ref}}]
           -\frac{1}{2} (ia \vert f_H \vert aa-ii)
   \nonumber \\
   & = & -\frac{1}{2} (ia \vert f_H \vert aa-ii) \, .
  \label{eq:theory.9.2}
\end{eqnarray}
Also
\begin{eqnarray}
  C & = & f_{i,a}^{\mbox{HF}}[\gamma] + (ai \vert f_H \vert aa - ii)
    \nonumber \\
    & = & f_{i,a}^{\mbox{HF}}[\gamma_{\mbox{ref}}] - \frac{1}{2}
    (ai \vert f_H \vert aa-ii) + (ai \vert f_H \vert aa -ii)
    \nonumber \\
    & = & +\frac{1}{2}(ai\vert f_H \vert aa-ii) \, .
  \label{eq:theory.9.3}
\end{eqnarray}
In fact,
\begin{equation}
  C = f_{i,a}^{\mbox{HF}}[\gamma_{i,{\bar i}}^{a,{\bar a}}] \, ,
  \label{eq:theory.10}
\end{equation}
because,
\begin{equation}
  \hat{\gamma}_{i,{\bar i}}^{a,{\bar a}} = \hat{\gamma}_{\mbox{ref}} + 
  \frac{1}{2} \left(\vert a \rangle \langle a \vert + 
  \vert \bar{a} \rangle \langle \bar{a} \vert \right) - 
  \frac{1}{2} \left( \vert i \rangle \langle i \vert 
  + \vert \bar{i} \rangle \langle \bar{i} \vert \right)
  \label{eq:theory.11}
\end{equation}
means that,
\begin{equation}
  C = f_{i,a}^{\mbox{HF}}[\gamma_{i,{\bar i}}^{a,{\bar a}}]
  = f_{i,a}^{\mbox{HF}}[\gamma] + \frac{1}{2} (ia\vert \vert aa-ii) 
  = \frac{1}{2} (ia\vert \vert aa-ii) \, ,
  \label{eq:theory.11.1}
\end{equation}
by reasoning analogous to that in Eq.~(\ref{eq:theory.9.2}).
Hence our choice of reference state yields
\begin{equation}
  D = -\frac{1}{2} ( ai \vert \vert aa - ii) = -C \, .
  \label{eq:theory.11.2}
\end{equation}

\begin{figure}
\begin{center}
\includegraphics[width=0.7\textwidth]{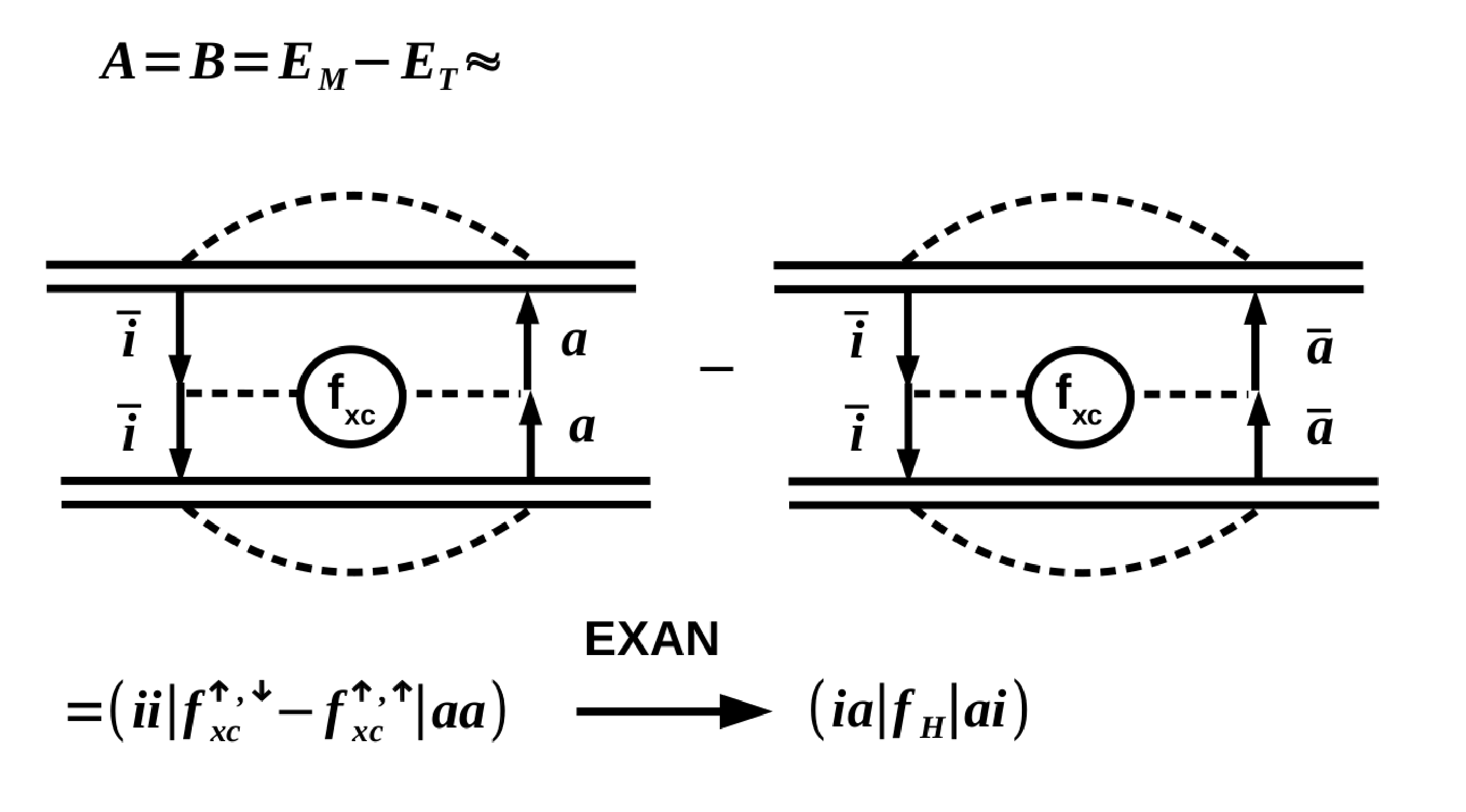}
\end{center}
\caption{
Diagrammatic MSM expressions for the off-diagonal CI matrix elements 
$A$ and $B$.
\label{fig:diagMSMa}
}
\end{figure}

\begin{figure}
\begin{center}
\includegraphics[width=0.4\textwidth]{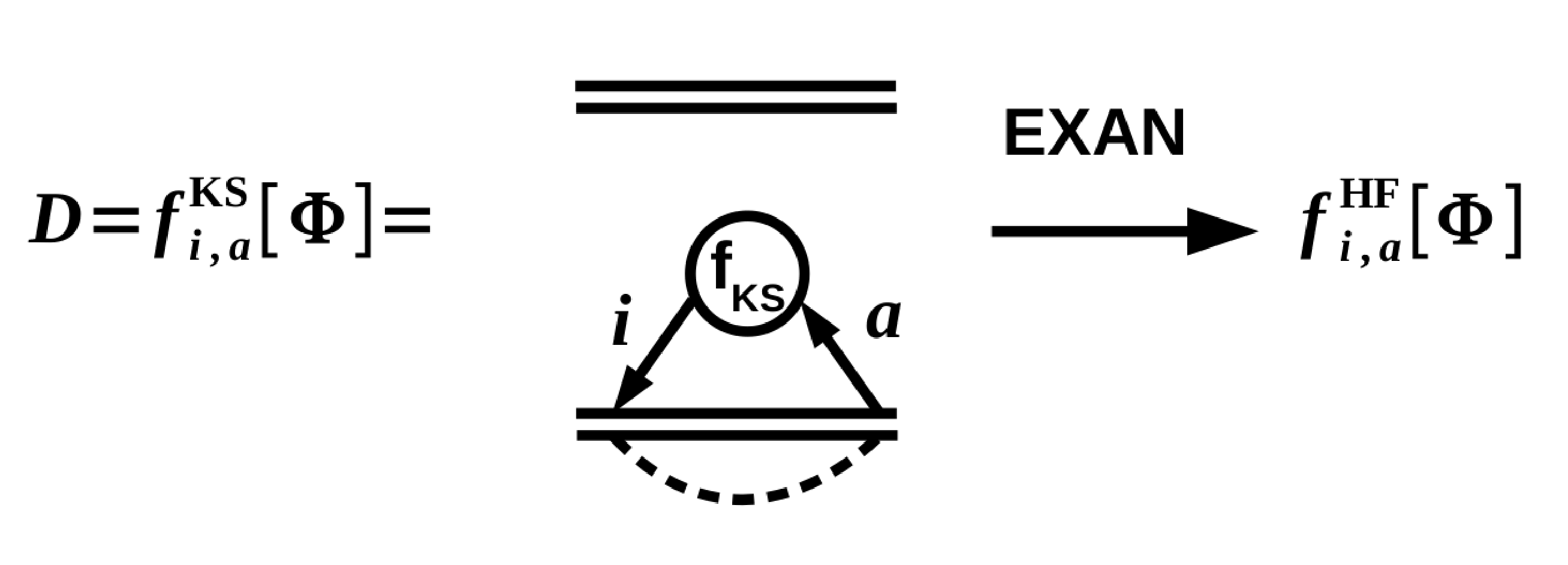}
\end{center}
\caption{
Diagrammatic MSM expressions for the off-diagonal CI matrix element 
$D$.
\label{fig:diagMSMb}
}
\end{figure}

\begin{figure}
\begin{center}
\includegraphics[width=0.7\textwidth]{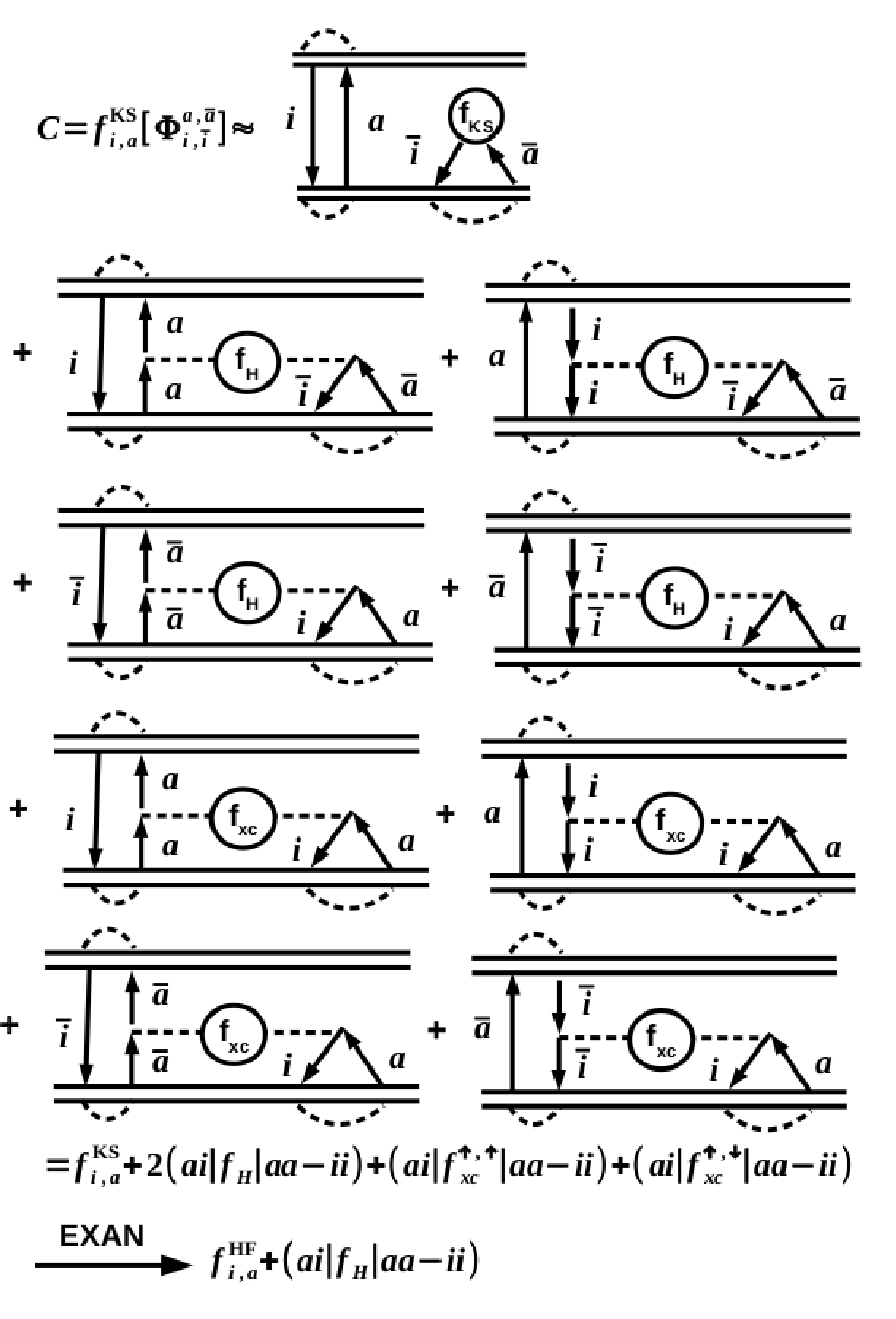}
\end{center}
\caption{
Diagrammatic MSM expressions for the off-diagonal CI matrix element 
$C$.
\label{fig:diagMSMc}
}
\end{figure}

For simplicity, we consider only pure DFAs --- i.e., those which depend
only upon the density (LDA depending upon $\rho_\uparrow$ and 
$\rho_\downarrow$) and its first (GGAs) and second (some meta GGAs) 
functionals, but no orbital dependence --- but the generalization to
hybrid DFAs is straightforward.  We will also assume that the
reference is invariant under exchange of spins so that 
$v_{xc}^\uparrow = v_{xc}^\downarrow$, 
$f_{xc}^{\uparrow,\uparrow} = f_{xc}^{\downarrow,\downarrow}$,
and 
$f_{xc}^{\uparrow,\downarrow} = f_{xc}^{\downarrow,\uparrow}$.
In the diagrammatic MSM-DFA method, we need to guess the form of 
the key matrix elements $A$, $B$, $C$, and $D$.  Our guesses are
shown in {\bf Fig.~\ref{fig:diagMSMa}} and in {\bf Fig.~\ref{fig:diagMSMb}} 
along with corresponding diagrams where appropriate.  The MSM
expression for the $A$ matrix element may be traced back to the original
Ziegler-Rauk-Baerends paper and comes from a spin symmetry analysis 
\cite{ZRB77}.  Our expression for the $B$ matrix element comes from
the argument in Article I that the CIMEs for $A$ and $B$ evaluate
to the same integral in WFT for real-valued orbitals.  We note that the
EXAN produces the MSM-HF expression.  The $D$ term shown in 
{\bf Fig.~\ref{fig:diagMSMb}} is also intuitive given the corresponding
CIME diagram, as long as we understand that   
\begin{equation}
  f_{i,a}^{\mbox{KS}}[\gamma] = \langle \Phi \vert \hat{f}_{\mbox{KS}} 
   [\gamma] a^\dagger i \vert \Phi \rangle \, .
  \label{eq:theory.12}
\end{equation}
To obtain the $C$ diagrams in {\bf Fig.~\ref{fig:diagMSMc}}, we need
to expand 
\begin{eqnarray}
   f_{i,a}^{\mbox{KS}}[\gamma_{i,{\bar i}}^{a,{\bar a}}]
   & = & f_{i,a}^{\mbox{KS}}[\gamma] \nonumber \\
   & + & \langle \Phi \vert v_H[\rho+2\rho_a-2\rho_i] 
      - v_H[\rho] \vert \Phi_i^a \rangle  \nonumber \\
   & + & \langle \Phi \vert 
    v_{xc}^\uparrow[\rho^\uparrow+\rho_a-\rho_i, \rho^\downarrow] 
    - v_{xc}^\uparrow \vert \Phi_i^a \rangle \nonumber \\
   & + & \langle \Phi \vert
    v_{xc}^\uparrow[\rho^\uparrow,\rho^\downarrow+\rho_a-\rho_i]
    - v_{xc}^\uparrow[\rho^\uparrow,\rho^\downarrow] \vert \Phi_i^a \rangle
  \nonumber \\
  & \approx & f_{i,a}^{\mbox{KS}}[\gamma] + 2 (ia \vert f_H \vert aa -ii )
  \nonumber \\
  & + & (ia \vert f_{xc}^{\uparrow,\uparrow}[\rho^\uparrow,\rho^\downarrow] \vert aa-ii)
  + (ia \vert f_{xc}^{\uparrow,\downarrow}[\rho^\uparrow,\rho^\downarrow] \vert aa-ii) \, .
  \label{eq:theory.13}
\end{eqnarray}

An explicit charge transfer correction is evident in Fig.~\ref{fig:diagMSMc}.
Note however that these diagrams have been chosen with $\Phi$ as
the choice of reference.  However, as discussed above, we have 
chosen to make a different choice of reference ($\rho=\rho_{\mbox{ref}}$
with half an electron of each spin promoted from $i$ to $a$) that actually leads, 
\begin{eqnarray}
  C & = & (ia \vert f_H \vert aa-ii) 
      + \langle i \vert v_{xc}^\uparrow[\rho^\uparrow+\frac{1}{2}(\rho_a-\rho_i),
      \rho^\downarrow+\frac{1}{2}(\rho_a-\rho_i)]
      - v_{xc}^\uparrow[\rho^\uparrow,\rho^\downarrow] \vert a \rangle \nonumber \\
    & \approx & +\frac{1}{2} (ai \vert 2f_H +f_{xc}^{\uparrow,\uparrow}
          +f_{xc}^{\uparrow,\downarrow} \vert aa-ii)
   \nonumber \\
  D & = &  -(ia \vert f_H \vert aa-ii) 
      + \langle i \vert v_{xc}^\uparrow[\rho^\uparrow-\frac{1}{2}(\rho_a-\rho_i),
      \rho^\downarrow-\frac{1}{2}(\rho_a-\rho_i)]
      - v_{xc}^\uparrow[\rho^\uparrow,\rho^\downarrow] \vert a \rangle \nonumber \\
    & \approx & -\frac{1}{2} (ai \vert 2f_H +f_{xc}^{\uparrow,\uparrow}
          +f_{xc}^{\uparrow,\downarrow} \vert aa-ii)
  \, ,
  \label{eq:theory.14}
\end{eqnarray}
so we expect $C \approx -D$, but not necessarily that $C = -D$.  

The advantage of using the dynamic correlation in SDET orbital operators 
is that we may calculate matrix elements
that would not otherwise have been possible to calculate within
the MSM approach.  They are also guaranteed, by construction, to
reduce to the usual CIME terms for the exact exchange part of 
hybrid functionals.  However more work has been done to 
characterize DFA total energies than orbital hamiltonian matrix elements.
Let us consider this a little further by using the exchange-only
LDA where
\begin{equation}
  v_x^\sigma[\rho^\sigma](\vec r) = -C_x (\rho^\sigma)^{1/3}(\vec r)
  \, ,
  \label{eq:theory.15}
\end{equation}
with
\begin{equation}
  C_x = \left( \frac{6}{\pi} \right)^{1/3} \, .
  \label{eq:theory.16}
\end{equation}
Then
\begin{equation}
  \frac{v_x^\sigma[\rho^\sigma \pm \Delta \rho^\sigma](\vec r)-v_x^\sigma[\rho^\sigma] (\vec r)}
  {v_x^\sigma[\rho^\sigma] (\vec r)}
  = \left( 1 \pm \frac{\Delta \rho^\sigma(\vec r)}{\rho^\sigma(\vec r)} \right)^{1/3}-1 \, .
  \label{eq:theory.17}
\end{equation}
The sign of the term depends upon the choice of ``+'' or ``-'' but the magnitude will
be smaller for ``+'' than for ``-''.  This implies that $C < -D$ because $C$ is calculated
using a higher value of the density.  This leaves us with four choices for calculating
$C$ and $D$ in our formalism, namely
\begin{enumerate}
  \item Unsymmetrized: Use Eq.~(\ref{eq:theory.14}) directly.
  \item Symmetrized: Replace $C$ and $D$ with 
        \begin{eqnarray}
           C' & = & \frac{C-D}{2} \nonumber \\
           D' & = & -C' \, .
           \label{eq:theory.18}
        \end{eqnarray}
  \item From $\Phi_D$: Replace $C$ and $D$ with
        \begin{eqnarray}
           C" & = & C \nonumber \\
           D" & = & -C \, .
           \label{eq:theory.19}
        \end{eqnarray}
  \item From $\Phi_0$: Replace $C$ and $D$ with
        \begin{eqnarray}
           C^{(3)} & = & -D \nonumber \\
           D^{(3)} & = & D \, .
           \label{eq:theory.20}
        \end{eqnarray}
\end{enumerate}
The reason for the second choice is that $C=-D$ is exact in WFT.  The third and fourth
choices require fewer calculations, which is important as our approach is, as yet, only
partially automated.  We will see how large the numerical differences between these
different choices are for the LDA in Sec.~\ref{sec:results}.
(For other DFAs, see the SI.)

%
%

\section{Computational Details}
\label{sec:details}

All calculations were done with the freely downloadable version of
{\sc deMon2k} ({\em densit\'e de Montr\'eal}, so called because it was
developed at the {\em Universit\'e de Montr\'eal}) \cite{deMon2k}.
The reader is directed to Article I for more information about the
computational details.  
Here we emphasize the steps needed to compute
the state energies using a program such as the {\sc Python} program
given in the SI.  At this stage in the development of our work, we
do most of the calculations by hand which requires the use of quite
a few files.  These files will have names such as {\tt 3p6HaLb.inp}
for one spin $\alpha$ ($\uparrow$) electron in the HOMO and one
spin $\beta$ ($\downarrow$) electron in the LUMO and a bond distance
of 3.6 (hence ``3p6'') bohr.  For the purpose of illustration, we 
will assume a bond length of 3.6 bohr in our file names, but it should 
be kept in mind that this is a parameter which is varied during our 
calculations.  Once sufficiently many calculations are done we use
a spread sheet (e.g., {\sc OpenOffice Calc}) to graph the results.

\paragraph{Step 0: Energy Zero}
Most PECs refer to an energy zero at infinite dissociation.  In the
case of LiH, we use the sum of the energies of the two neutral atoms
calculated using the same functional and basis set. This provides
us with the energy {\tt EZ}.  Note that our PECs will {\em not} dissociate
to exactly this energy because of our choice of reference state.  Hence
additional curve shifting may be desirable to redefine the energy zero.

\paragraph{Step 1: Reference State}
The input file is {\tt 3p6ref.inp}.
The keyword {\tt MOMODIFY} allows us to carry out a spin-unrestricted
calculation with half an electron of each spin in the HOMO and in the LUMO.
In the specific case of LiH, we used
\begin{verbatim}
MOMODIFY 2 2
2 0.5
3 0.5
2 0.5
3 0.5
\end{verbatim}
Running the calculation provides us with a restart file 
({\tt 3p5ref.rst} in this case) for our other calculations
and with the MO coefficients which are placed in the matrix {\tt C}
in the {\sc Python} program.  The program also prints out the 
\marginpar{\color{blue} AO}
Kohn-Sham matrix in the atomic orbital (AO) basis set.  Although this
is not needed for our calculations, it is reassuring that the {\sc Python}
program transforms this Kohn-Sham matrix to the MO basis so that we may
verify that it is indeed diagonal.

\paragraph{Steps 2-5: Diagonal Energies}
The input files are prepared for the ground state ({\tt 3p6HaHb.inp}), 
triplet ({\tt 3p6HaLa.inp}), mixed symmetry ({\tt 3p6HaLb.inp}),
and doubly excited determinants ({\tt 3p6LaLb.inp}) in order to
calculate the corresponding energies {\tt EG}, {\tt ET}, {\tt EM},
and {\tt ED}.  Each needs to include the key line
\begin{verbatim}
SCFTYPE UKS MAX=0
\end{verbatim}
that ensures that the reference MOs are used to calculate the 
energy {\tt ET} without any SCF optimization with the reference MOs and
the occupation numbers determined by the {\sc MOMODIFY} keyword.
For example, the triplet energy is calculated using 
\begin{verbatim}
MOMODIFY 2 2
2 1.0
3 1.0
2 0.0
3 0.0
\end{verbatim}
which tells us that 2 spin $\alpha$ and 2 spin $\beta$ orbital occupations
are going to be modified.  In this case the HOMO is the second MO and the
LUMO is the third MO. It is also important that the input files contain
the keyword
\begin{verbatim}
GUESS RESTART
\end{verbatim}
which tells the program to read the MOs from the restart file.
The keyword line
\begin{verbatim}
PRINT MOS KS
\end{verbatim}
is also needed for the ground and doubly-excited determinants.
We will denote the MO coefficient matrix for the different calculations
as {\tt Cref}, {\tt CG}, and {\tt CD} for, respectively the reference
calculation the $\Phi_0$ calculation and the $\Phi_D$ calculation.
Similarly the Kohn-Sham matrices in the AO basis sets are {\tt Fref},
{\tt FG}, and {\tt FD}.
The program is run by copying {\tt 3p6ref.rst} to {\tt 3p6HaLb.rst}
(still for the triplet case) and then using a run shell (not described
here but see Refs.~\cite{C22,OPEC23})
\begin{verbatim}
  ./run.csh 3p6HaLb
\end{verbatim}
and similarly for the other files.  Energies are read from {\tt 3p6HaLb.out}
and other output files to assign the variables {\tt EG}, {\tt ET}, {\tt EM},
and {\tt ED} in the {\sc Python} program.  

\paragraph{Steps 6-7: Kohn-Sham Matrices}
We also need the Kohn-Sham matrix in the AO basis set for the ground and 
doubly-excited states.  These are obtained analogously to what is done in
steps 2-5. However we are faced with the problem that {\sc deMon2k} will
not output a new Kohn-Sham matrix unless we use
\begin{verbatim}
SCFTYPE UKS MAX=1
\end{verbatim}
to force the program to construct and output a Kohn-Sham matrix.  As it
is important that this Kohn-Sham matrix is constructed from the original
orbitals and not updated, we need to use a trick---namely the {\tt MIXING} 
keyword 
\begin{verbatim}
GUESS RESTART
MIXING +0.0000000000
\end{verbatim}
This prevents the program from updating the initial guess.
Note that {\sc deMon2k} was never designed for this type of calculation
and will give an error statement.  Nevertheless, we obtain the desired
correct Kohn-Sham matrix.
We use the file name {\tt 3p6HaHb3.inp} to distinguish from {\tt 3p6HaHb.inp}.
The resultant Kohn-Sham matrices in the AO representation are stored in 
the {\tt FG} and {\tt FD} matrices in the {\sc Python} program.  
However the real Kohn-Sham matrices are (for a reason linked to how 
{\sc deMon2k} treated restricted open-shell Kohn-Sham calculations) 
twice this value.  Hence we need to multiply by two ({\tt FG2=2*FG} 
and {\tt FD2=2*FD}).
By {\tt FGref} and {\tt FDref} we mean the Kohn-Sham matrices for
$\Phi_0$ and for $\Phi_D$ in the reference MO basis set which are
constructed as {\tt FGref} = {\tt Cref}$^\dagger$ {\tt FG2} {\tt Cref} 
and {\tt FDref} = {\tt Cref}$^\dagger$ {\tt FD2} {\tt Cref}.

As one of the most time consuming parts of our procedure is
reformating matrices from the {\sc deMon2k} output format to {\sc Python}
format, additional auxiliary {\sc Python} programs were written to
extract these matrices from the {\sc deMon2k} output and format
them for inclusion in our principle {\sc Python} program.
These helper programs are also included in the SI.


\paragraph{Basis Sets}
\marginpar{\color{blue} GTO}
{\sc deMon2k} uses gaussian-type orbital (GTO) basis sets.
As in Article I, we use the {\tt DZVP} orbital basis set, but we have chosen
to upgrade to the larger {\tt GEN-A3*} auxiliary basis set.

\paragraph{Functionals}
The {\tt FOCK} keyword in {\sc deMon2k} provides an auxiliary-function 
calculation of HF exchange so that we also have access to a good approximation 
to Hartree-Fock and to hybrid functionals.  
\marginpar{\color{blue} LDA}
For the local density approximation (LDA), we use
the Vosko-Wilk-Nusair (VWN) parameterization \cite{VWN80}
of Ceperley and Alder's quantum Monte Carlo results \cite{CA80} for
the homogeneous electron gas
[confusingly denoted VWN5 in the popular {\em Gaussian} program where
VWN is (mistakenly) used to designate the parameterization of random-phase
\marginpar{\color{blue} RPA, GGA, PW91, B88, B3LYP}
approximation (RPA) results reported in the Vosko-Wilk-Nusair article 
\cite{VWN80}].  We have also chosen one generalized gradient approximation 
(GGA), namely the Perdew-Wang 1991 (PW91) functional \cite{PCV+92}.
Finally we chose one global hybrid functional, namely the three-parameter
Becke exchange plus Lee-Yang-Parr (B3LYP) functional \cite{SDCF94}.
The B3LYP functional implemented in {\sc deMon2k} is the same as that
implemented in {\sc Gaussian}---notably in using the VWN parameterization
of the RPA results, rather than their parameterization of the 
Ceperley-Alder quantum Monte Carlo results---except that the Becke's 1988
exchange functional (B88) \cite{B88} has been modified to satisfy the 
Lieb-Oxford inequality \cite{L79a,LO81,CH99,LLS22} and, of course, that 
{\sc deMon2k} uses an auxiliary-function approximation to HF exchange.


\paragraph{Relaxed and BS Results}
This article is concerned with LiH PECs calculated without symmetry
breaking using a novel approach.  This approach uses the unrelaxed
orbitals from a reference state, rather than the usual relaxed ground
state orbitals.  As such, we must expect that the well in the ground state 
PEC will be predicted to be a bit too shallow, which is to say that binding
energies are expected to be underestimated.  In order to provide a reality 
check,
{\bf Table~\ref{tab:MSM_BE}} provides optimized ground-state bond lengths 
and binding energies obtained in the usual way (denoted as ``relaxed''), 
using the same functionals, 
basis sets, auxiliary basis sets, and grids.  In particular, binding energies
are calculated by taking the difference between the calculated energy at the 
optimized molecular geometry and the sum of the calculated energies of the two 
neutral atoms.  

The BS results in Fig.~\ref{fig:normalLDA} are difficult to converge at bond
distances exceeding (and especially near) the CFP. to do so, we began at 10.0
{\AA} by restarting our multiplicity 1 calculation from a multiplicity 3
(i.e., triplet) calculation.  Various techniques were used to converge
\marginpar{\color{blue} SCF, OMA}
the self-consistent field (SCF) calculation, but the most effective method
was found to be the optimal mixing algorithm (OMA) \cite{C01}.
We then gradually shortened the LiH bond, using OMA and restarts from the 
previous geometry.  Symmetry is automatically restored at the 
CFP.

%
%

\section{Results and Analysis}
\label{sec:results}

The results obtained with the various DFAs are qualitatively---even
quantitatively---very similar.  Of course, this is exactly what we
want---namely a theory whose results are insensitive to the choice of
functional, as any robust DFT approach should be.  So, except when we 
explicitly want to compare DFAs, we will focus on the LDA (which we
also refer to as VWN after the name of the {\sc deMon2k} program option).  
Additional graphs with other DFAs may be found in the SI.

\begin{figure}
\begin{center}
\includegraphics[width=0.70\textwidth]{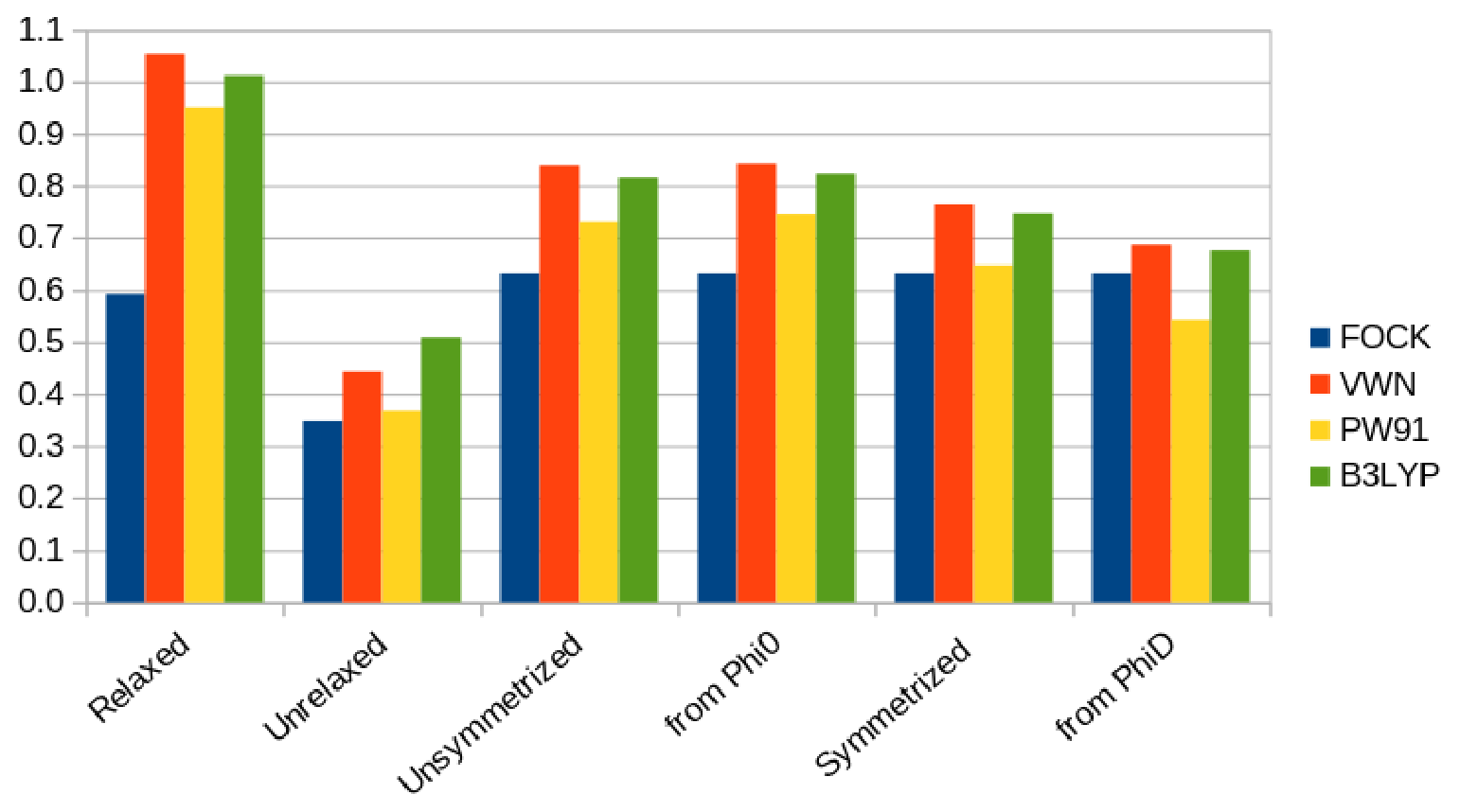}
\end{center}
\caption{
Ratio of binding energies to the EXACT binding energy at various
levels of approximation.
\label{fig:BindingEnergyRatios}
}
\end{figure}
Binding energies turn out to be much more sensitive to our choice of
method than are bond distances.
{\bf Figure~\ref{fig:BindingEnergyRatios}} provides a graphical summary
of the binding energy results from Table~\ref{tab:MSM_BE}.

The {\em relaxed} result is a normal GS calculation.  At this level HF recovers
only about 60\% of the EXACT binding energy while all three DFAs do much
better.  VWN is known to overestimate binding energies, but this overestimation
is not very large in the present case. PW91 over corrects the binding energy,
and B3LYP does quite well.  

The {\em unrelaxed} result is the same calculation
but using the orbitals obtained from the ensemble calculation.  It follows
from the variational principle that the relaxed binding energy will be an 
upper bound to the unrelaxed binding energy.  HF and the
DFAs are much more similar at this level and all seriously underestimate
the binding energy.  

\begin{figure}
\begin{center}
\includegraphics[width=0.8\textwidth]{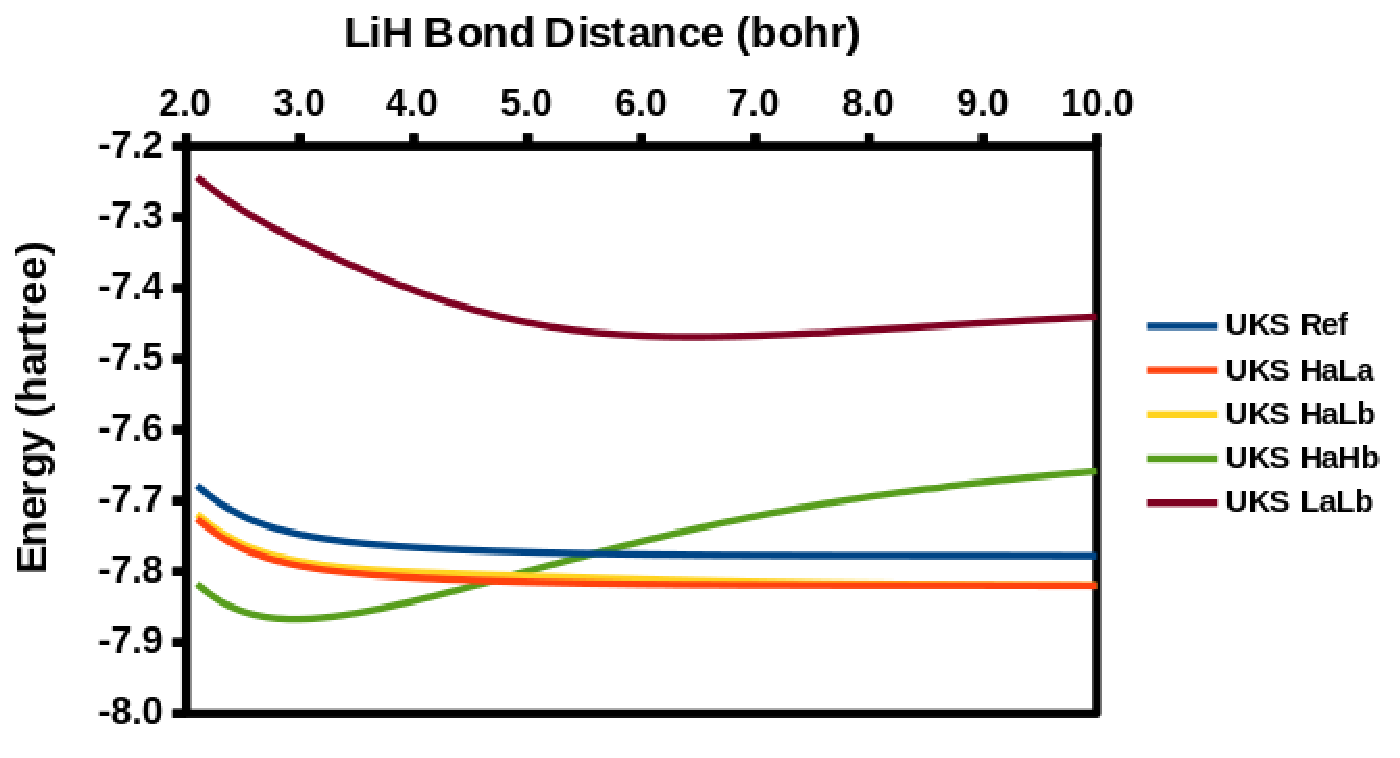}
\end{center}
\caption{
LDA SDET reference state energy and other SDET energies calculated with
the resultant MOs but different occupation numbers.
\label{fig:LiH_VWN_SDEs}
}
\end{figure}

We wish to illustrate a typical calculation by presenting the results
obtained at various steps of the LDA calculation.  The MOs and their
energies have already been presented in Fig.~\ref{fig:LiH_LDA_MOs}
and discussed in the introduction (Sec.~\ref{sec:intro}).  These orbitals
are used to construct the SDET states used in our calculations and
whose energies are shown as a function of bond distance in 
{\bf Fig.~\ref{fig:LiH_VWN_SDEs}}.  The triplet and mixed-symmetry
SDET calculations typically give very similar energies on the scale
of this graph.

\begin{figure}
\begin{center}
\includegraphics[width=\textwidth]{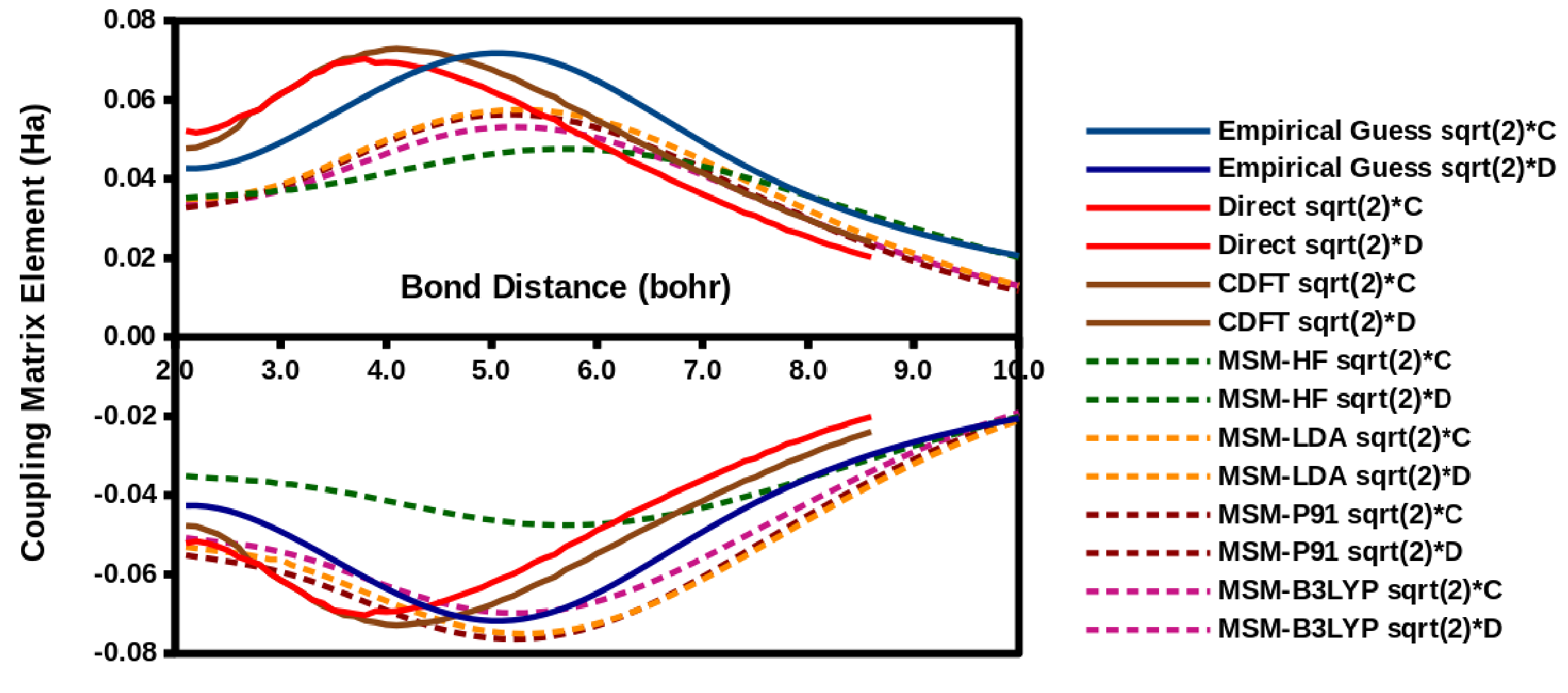}
\end{center}
\caption{
Graph of the $\sqrt{2}C>0$ and $\sqrt{2}D<0$ matrix elements obtained
from MSM-DFT.  The empirical curves are a gaussian guess at the coupling
matrix element made in the SI of Article I.  The ``direct'' and ``CDFT'' curves
were obtained from Fig.~30 of Ref.~\cite{KKV12} after digitization
with {\sc WebPlotDigitizer} \cite{WebPlotDigitizer} and conversion to the
format shown here.  [The details of how these last two curves were
calculated have not been published but we have confirmed the essential
points of their calculations with one of the authors (TV) in order to
be able to make the present comparison.]
\label{fig:coupling}
}
\end{figure}

The next step is to calculate the $C$ and $D$ coupling matrix elements.
{\bf Figure~\ref{fig:coupling}} is perhaps {\em the most important result of
this article} as it shows how well we can calculate the coupling matrix
elements $C$ and $D$ within the proposed choice of approximations.  
We have two reality
checks available to us---namely a gaussian coupling matrix estimate from
the SI of Article I and coupling matrix elements from Fig.~30 of 
Ref.~\cite{KKV12}.  As will become clearer below, these give us 
$-\sqrt{2}D$ but we have {\em assumed} 
$\sqrt{2} C = -\sqrt{2} D$ for the purpose of graphing these reality checks.

As expected [Eq.~(\ref{eq:theory.11})],
MSM-HF coupling matrix elements in Fig.~\ref{fig:coupling} 
rigorously satisfy the condition that $C=-D$.  Compared with our 
reality checks, the magnitude of the MSM-HF coupling elements is
significantly underestimated.  However, in general $C<-D$ in
MSM-DFT as discussed after Eq.~(\ref{eq:theory.17}).  
The $C$ and $D$ MSM-LDA and MSM-PW91 coupling elements are, respectively,
 very similar to each other.  Not surprisingly, as B3LYP contains a 
portion of HF exchange, the MSM-B3LYP coupling matrix elements are slightly
shifted towards the MSM-HF result.  Focusing now on the $\sqrt{2}D<0$
curves, we see that MSM-DFT is giving the right magnitude for this coupling
matrix element with the MSM-B3LYP coupling element being arguably a bit
better than for those obtained from MSM-LDA and MSM-PW91 calculations.
On the other hand, the MSM-DFT $C$ coupling element is closer in magnitude
to that obtained for MSM-HF.

Referring back to Table~\ref{tab:MSM_BE} and Fig.~\ref{fig:BindingEnergyRatios},we see that the full (unsymmetrized) calculation, including the coupling 
matrix elements $C$ and $D$, increases the binding energy again compared
to the unrelaxed calculation.  It is important to realize that there
are two reasons for this happening---namely (i) the coupling matrix elements 
allow mixing of the $i$ and $a$ orbitals which simulates relaxation to some 
extent, thus increasing the binding energy and (ii) the coupling matrix
elements add in additional strong correlation. What does {\em not} happen
to any significant amount is the addition of more dynamic correlation as
the $3 \times 3$ CI matrix is simply too small to contribute 
more than an infinitessimal
amount of dynamic correlation beyond that which is already present in the basic
DFA used.

\begin{figure}
\begin{center}
\includegraphics[width=\textwidth]{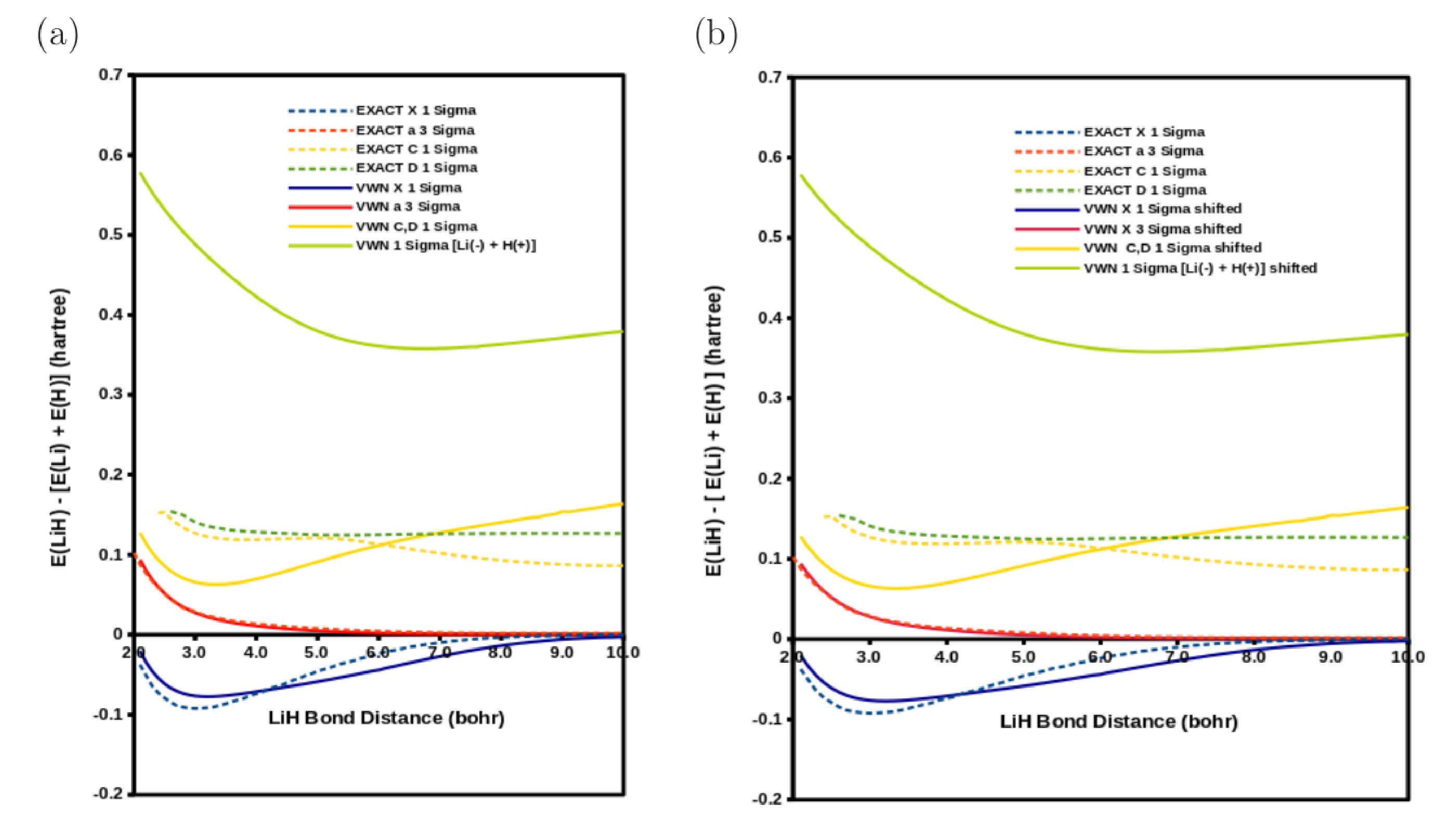} 
\end{center}
\caption{
Comparison of EXACT and MSM-LDA PECs calculated using the {\em unsymmetrized}
matrix elements: (a) energy zero calculated as the
sum of HF energies calculated for Li$^+$ and H:$^-$, (b) shifted to 
that the $a \,^3\Sigma$ curve goes to zero at $R = \mbox{10.0 bohr}$.
\label{fig:LiH_LDA_PECs}
}
\end{figure}
Although our interest is mainly in the GS PEC, it is good to have a look
at all the PECs. {\bf Figure~\ref{fig:LiH_LDA_PECs}} shows the four
MSM-LDA curves compared with the EXACT results.  The energy in part (b) 
is shifted so that the triplet energy goes to zero at $R = \mbox{10.0 bohr}$
while part (a) is relative to the energy of the neutral atoms calculated
at the same level.  For MSM-LDA, MSM-PW91, and MSM-B3LYP, there is no
noticable difference between graphs (a) and (b).  In MSM-HF however, 
the triplet curve in graph (a) is noticably greater than zero at 
$R = \mbox{10.0 bohr}$ due, presumably, to the neglect of relaxation 
effects when using the ensemble reference.  While the MSM-DFT $a \,^3\Sigma$
curves are in excellent agreement with the corresponding EXACT curves,
the MSM-DFT $C,D \,^1\Sigma$ curves are significantly underestimated.
In contrast, the MSM-DFT GS curve is in the right energy range and separates
correctly to the sum of the energies of the two neutral atoms.  In this sense,
we have succeeded in obtaining a correct description of the 
[Li$^+$   H:$^-$]/[Li$\uparrow$   H$\downarrow$ $\leftrightarrow$ 
Li$\downarrow$   H$\uparrow$] avoided crossing.  There is no EXACT
PEC for the $^1 \Sigma$[Li:$^-$   H$^+$] state.

\begin{figure}
\begin{center}
\includegraphics[width=0.8\textwidth]{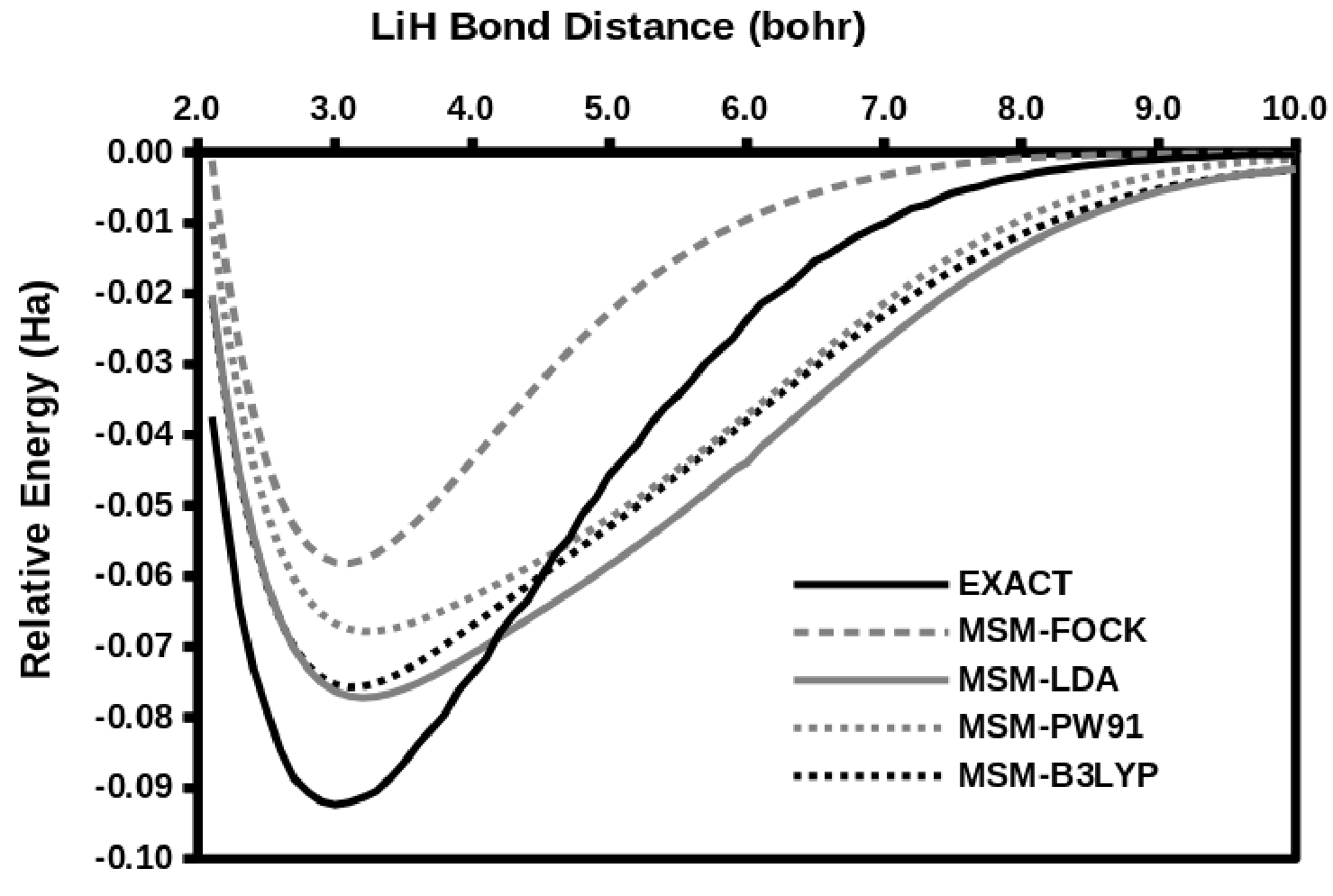}
\end{center}
\caption{
Comparison of the exact $X\, ^1\Sigma$ and diagrammatic MSM
PECs obtained with different DFAs using the {\em unsymmetrized}
coupling matrix element.
\label{fig:GroundStatePEC3}
}
\end{figure}
Let us return now to the GS curve which is, after all, our primary interest.
PECs obtained with the diagrammatic MSM are compared with the EXACT PECs
in {\bf Fig.~\ref{fig:GroundStatePEC3}}.  We see that all the 
ensemble-referenced MSM PECs are underbound compared with the EXACT
PEC.  Also the MSM-HF PEC is the most shallowly bound with the three
MSM-DFT PECs being more similar.  Notice that we do not get quantitative
accuracy for the binding energy because of the choice of reference orbitals.
A subtler problem is that the {\em shape} of the MSM-HF PEC is qualitatively
in better agreement with the shape of the EXACT PEC than is the case for
the three MSM-DFT PECs.  In particular, the three MSM-DFT PECs are {\em
higher} in energy than the EXACT PEC near the equilibrium geometry but fall
{\em lower} in energy than the EXACT PEC as LiH dissociates.  So this model
is clearly a step in the right direction, but it not yet the quantitative
method that we would like.

\begin{figure}
\begin{center}
\includegraphics[width=0.6\textwidth]{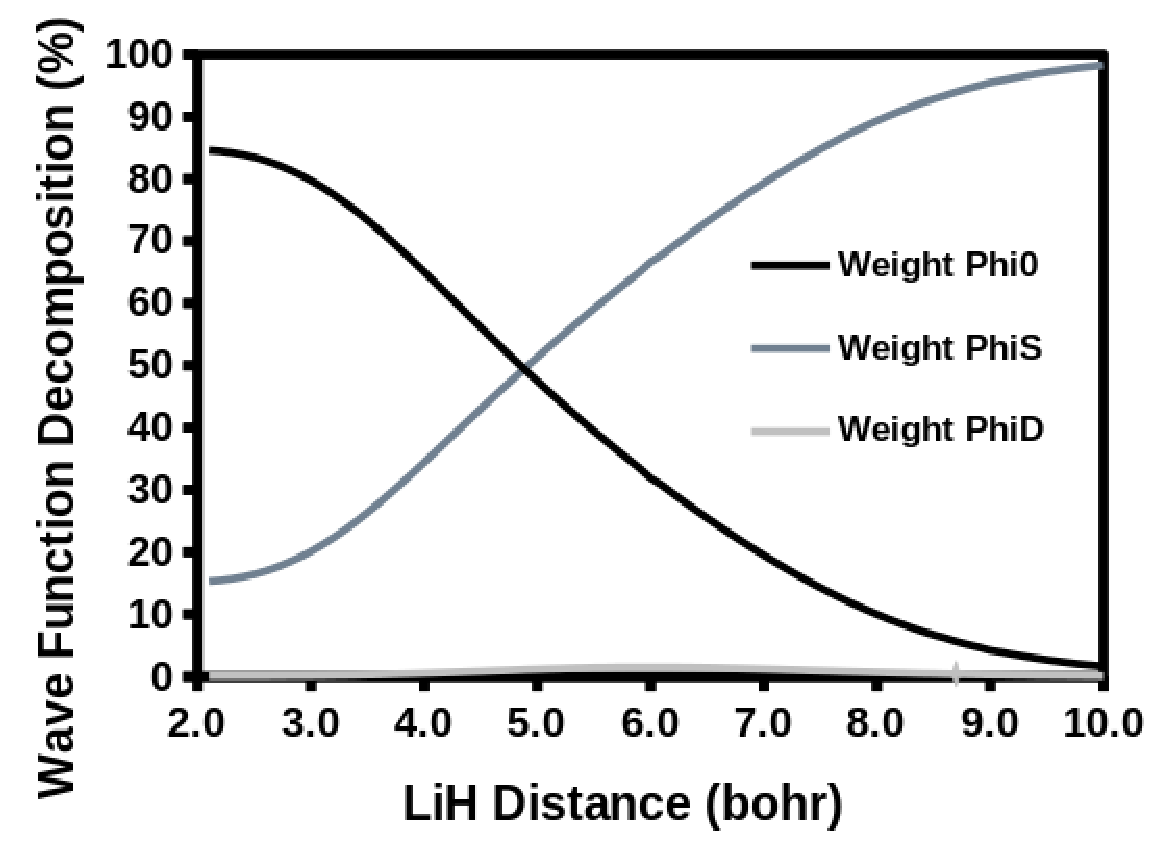}
\end{center}
\caption{
Decomposition of the diagrammatic
MSM-LDA CI wave function calculated with {\em unsymmetrized} matrix
elements as a function of LiH bond length:
Weight Phi0, $\vert H, \bar{H}\vert$;
Weight PhiS, $(1/\sqrt{2})(\vert H, \bar{L} \vert + \vert L, \bar{H} \vert )$;
Weight PhiD, $\vert L, \bar{L} \vert$.  The Phi0 and PhiS weights cross
at about 4.9 bohr.
\label{fig:LDA_popanal}
}
\end{figure}
One of the more beautiful features of using the ensemble reference is
the symmetrical way that the HOMO (H, $i$) and LUMO (L, $a$) are treated 
symmetrically in this formalism.  This comes out particularly strongly
when looking at the MSM-DFT wave-function decomposition.  This is shown for the 
MSM-LDA wave-function in {\bf Fig.~\ref{fig:LDA_popanal}}.  We see that the 
doubly excited state plays almost no role.  Rather, the GS and open-shell 
singlet states mix strongly over the entire range of bond distances studied
with equal contributions somewhere around 5 bohr.  This is strong evidence
that the $D$ coupling matrix element is {\em much} more important than the 
$C$ matrix element.  This is the case also for our MSM-HF, MSM-PW91, and 
MSM-B3LYP calculations.

\begin{figure}
\begin{center}
\includegraphics[width=0.65\textwidth]{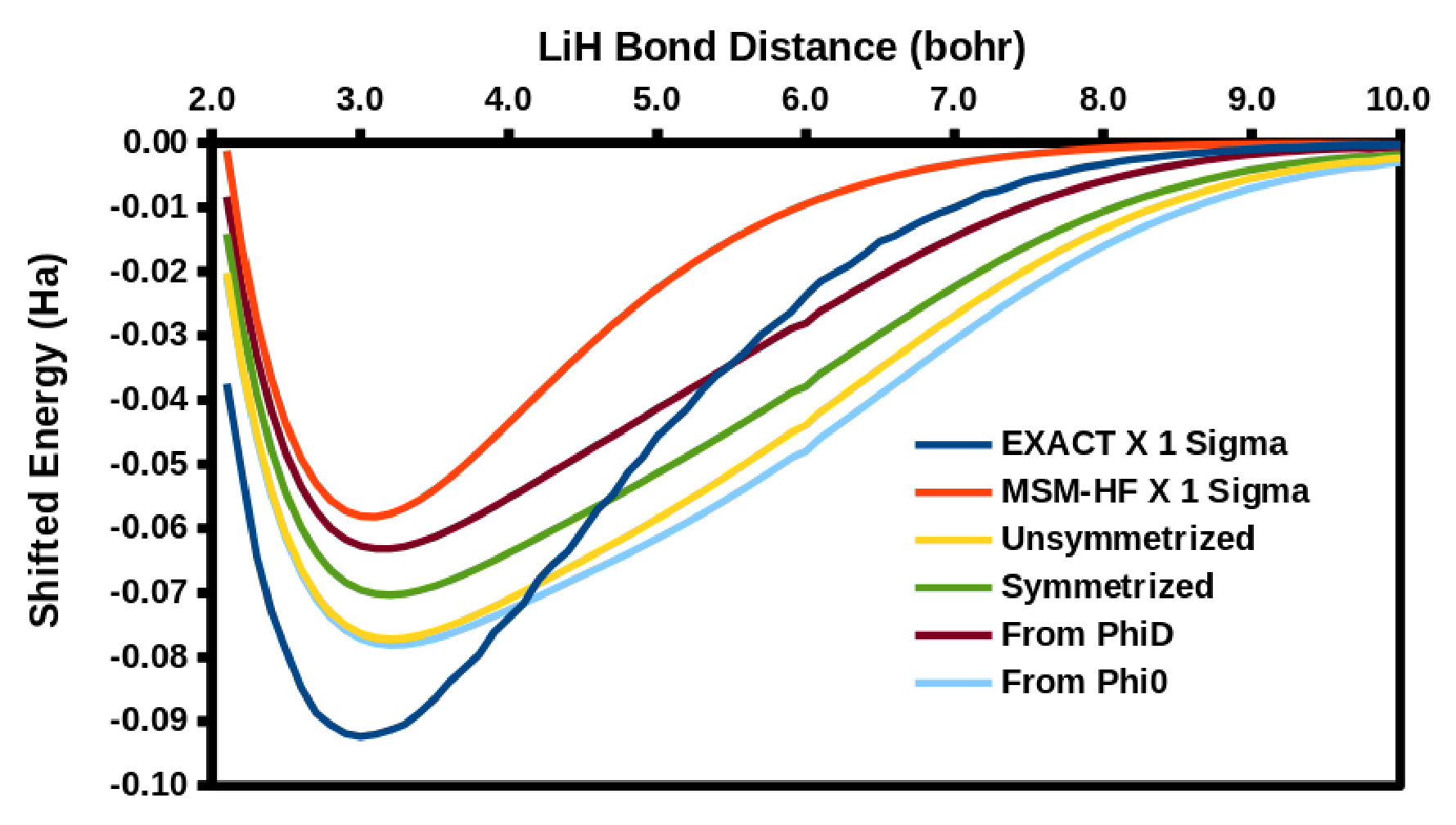}
\end{center}
\caption{
Comparison of the EXACT $X\, ^1\Sigma$, MSM-HF $X\, ^1\Sigma$,
and diagrammatic MSM-LDA PECs for different choices of the $C$ and $D$
matrix elements:
{\em unsymmetrized} [$C$ and $D$ in Eq.~(\ref{eq:theory.14})],
{\em symmetrized} [$C'$ and $D'$ in Eq.~(\ref{eq:theory.18})],
{\em from $\Phi_D$} [$C"$ and $D"$ in Eq.~(\ref{eq:theory.19})],
and {\em from $\Phi_0$} [$C^{(3)}$ and $D^{(3)}$ from
Eq.~(\ref{eq:theory.20})].
\label{fig:VWNcompare}
}
\end{figure}
Last, but not least, it remains to compare the four different choices
for calculating the $C$ and $D$ coupling matrix elements listed at the
end of Sec.~\ref{sec:theory}.  These choices are compared for the 
MSM-LDA GS PEC in {\bf Fig.~\ref{fig:VWNcompare}}.  The {\em unsymmetrized}
method would seem to be the best choice from the point of view of 
bringing the binding energy as close as possible to that of the EXACT
PEC, and this is the one that we recommend.  However $C < -D$ which is 
different from the relation $C=-D$ which may be shown to hold exactly in 
MSM-HF. Hence a {\em symmetrized} method was also proposed.  Some may also 
wish for the coupling element to be closer in magnitude to that in MSM-HF.  
For those who wish this, the {\em from $\Phi_D$} choice is recommended 
which replaces $D$ with $-C$.  Or we could replace $C$ with $-D$ in the 
{\em from $\Phi_0$} method.  As seen in the MSM-DFT wave-function 
decomposition, it is really only the $C$ matrix element that matters as 
this is what couples the GS and open-shell singlet states.  This explains 
why the {\em from $\Phi_0$} choice PEC is close to the full 
{\em unsymmetrized} choice PEC.  
Evidently, the $2 \times 2$ matrix,
\begin{equation}
  {\bf H}_S = \left[
  \begin{array}{cc} E_0 & \sqrt{2} D \\
                  \sqrt{2}D & E_M+A \end{array}
  \right] \, ,
  \label{eq:results}
\end{equation}
would be adequate for most purposes when calculating the GS with the 
unsymmetrized method.
Not too surprisingly, the {\em symmetrized} 
choice gives a PEC which is in between the {\em from $\Phi_0$} and 
{\em from $\Phi_D$} PECs. 

At the risk of redundancy, we end this section by emphasizing that we prefer
the {\em unsymmetrized} choice as this gives the largest binding energy.
The avoided crossing is described adequately enough that the MSM-DFT GS
PEC separates to the energies of the separated neutral atoms.  
{\em Most remarkably, the magnitude of the $D$ coupling matrix element is
excellent.}
The method
is not yet quantitative but is certainly a promising step in the direction
of a parameter-free MDET DFT.

%
%

\section{Concluding Discussion}
\label{sec:conclude}

Lithium hydride, LiH, is an ideal testing ground for methods
treating strong nondynamic correlation in DFT.  It is small enough
that this entire study could be carried out with a freely downloadable
version of the {\sc deMon2k} program on our laptop computers.  Yet
LiH presents several challenges for DFT.

As we have noted, normal spin-unrestricted symmetry-broken DODS 
calculations give a reasonable-looking GS PEC.  Nevertheless, the 
PNDD leads to dissociation into fractionally charged atoms which is,
of course, physically incorrect.  Furthermore TD-DFT calculations
show a triplet energy which goes to zero at the CFP and becomes 
imaginary at larger bond distances with associated degredation of
the quality of the excited-state singlet PECs.  For all of these
reasons, we are searching for a parameter-free SODS MDET DFT which, when
combined with response theory, may provide a convenient approach to
problems involving strong correlation.

In Article {\bf I}, we proposed a diagrammatic approach as an aid to 
finding parallels between WFT and DFT that might otherwise be missed.
The idea was to begin with the FPP of MSM-DFT that DFAs are designed
to do a good job in describing dynamic correlation but not strong
correlation.  One way of thinking about the original 
Ziegler-Rauk-Baerends-Daul MSM is that it uses spin and spatial symmetry 
to find off-diagonal matrix elements in a small WFT CI matrix whose
diagonal elements include dynamic correlation via SDET DFT calculations.
That is, the traditional MSM uses symmetry to include static (i.e., 
degeneracy) correlation which is one type of strong correlation missed
by most DFAs.  Another type of strong correlation missed by most DFAs
is nondynamic (i.e., quasidegeneracy) correlation which is typical of 
making and breaking bonds.  The goal of the diagrammatic MSM-DFT approach
is to bypass spin and spatial symmetry to make guesses for other matrix
elements in a small CI matrix.  These are intended to be educated guesses, 
inspired by the structure of the corresponding WFT CI matrix and involving
only SDET DFT calculations and orbitals from, in our case, an ensemble
reference.  In addition, we must recover the WFT expressions
for matrix elements when we apply the EXAN.  Article I illustrated these
principles for H$_2$ and O$_2$.  A first attempt was also made with LiH
in the SI of Article I but something was missing!

The primary contribution of the present paper is to show how to calculate 
the missing key coupling matrix elements responsible for the 
[Li$^+$  H:$^-$]/[Li$\uparrow$  H$\downarrow$ $\leftrightarrow$ 
Li$\downarrow$  H$\uparrow$] avoided crossing needed for a proper
dissociation of the GS PEC within the TOTEM.  We do this by introducing 
a new idea---namely, by calculating off-diagonal matrix elements of the 
SDET Kohn-Sham orbital hamiltonians.  The magnitude of the coupling 
matrix elements obtained in this fashion are quite encouraging mainly, 
we think, because of the use of an ensemble reference that strikes a 
balance between the ground and open-shell singlet excited state.  Indeed 
the GS PEC does dissociate correctly and the general nature of the 
diagrammatic MSM-DFT coupling elements and calculated PECs is relatively
insensitive to the choice of DFA as would be hoped for a correctly formulated
parameter-free MSM-DFT. 

These results are promising (and not without a certain formal symmetry
and beauty) but will need further development in order to become the 
quantitative method that we are seeking.  On-going work suggests that
we can indeed improve the quality of the results by using a different 
reference state or, perhaps, by using multiple reference states.  More
work will be needed to see just how far we can go with these ideas.

%
%

\section*{Acknowledgements}
\label{sec:thanks}


The authors are grateful to the series of African Schools on Electronic 
Structure Methods and Applications (ASESMA) for encouraging pan-African
collaborations such as our own, teaching high-level courses with plenty
of discovery-based learning, and generally improving the research 
environment for theoretical solid-state and chemical physics in Africa.
We also thank Prof.\ {Van~Voorhis} for answering questions that we had 
about Fig.~30 of Ref.~\cite{KKV12} and for supplying us with a copy of 
Ref.~\cite{K12}.

\section*{Electronic Supplementary Information}
\label{sec:suppl}

The SI for this article contains:
\begin{enumerate}
\item Author Contributions
\item Particle Number Derivative Discontinuity
\item Python Programs
\item Results for Other Functionals
\end{enumerate}


\end{document}